\documentclass[11pt]{article}
\pdfoutput=1
\usepackage{jheppub}

\usepackage{ccaption}
  \captionnamefont{\bfseries}
  \captiontitlefont{\small\sffamily}
  \captiondelim{: }
  \hangcaption

\def\bulkJ{{\tilde J}}

\def\bulk{{\cal M}} 
\def\bdy{{\cal B}}

\def\regA{{\cal A}} 
\def\regAc{{\cal A}^c} 
\def\rhoA{{\rho_{\regA}}} 
 
\def\entsurf{\partial \regA} 
\def\domdA{D[\regA]}

\def\extr{{\cal E}_\regA} 
 
\def\homsurfA{{\cal R}_{\regA}} 
\def\homsurfAc{{\cal R}_{\regA}^c}

\def\EWA{{\cal W}_{\cal E}[\regA]}

\def\tE{t_\text{\tiny E}}
\newcommand{\fixM}{\mathbf{e}}

\newcommand\tauA{\tau_{_\regA}}
\newcommand{\CSA}{\Sigma_{_t}}
\newcommand{\bulkCS}{{\tilde \Sigma}_{_t}}

\def\tr{\text{Tr}}
\def\Zn{\mathbb{Z}_q}
\def\({\left(}
\def\){\right)}

\def\ket#1{\mid\!\! #1\rangle} 
\def\bra#1{\langle \, #1 \! \mid\! \ }

\definecolor{rust}{rgb}{0.8,0.2,0.2} 
\definecolor{purple}{rgb}{0.8,0.1,0.9} 
\definecolor{olivegreen}{rgb}{0,0.52,0.17}

\title{Deriving covariant holographic entanglement}

\author[a]{Xi Dong}
\author[b]{\!\!, Aitor Lewkowycz}
\author[c]{\!\!, Mukund Rangamani}
\affiliation[\,a]{School of Natural Sciences, Institute for Advanced Study, Princeton, NJ 08540, USA}
\affiliation[\,b]{Jadwin Hall, Princeton University,  Princeton, NJ 08544, USA}
\affiliation[\,c]{Center for Quantum Mathematics and Physics (QMAP)  \\
Department of Physics, University of California, Davis, CA 95616 USA}

\emailAdd{xidong@ias.edu}
\emailAdd{aitor@princeton.edu}
\emailAdd{mukund@physics.ucdavis.edu}

\abstract{
We provide a gravitational argument in favour of  the covariant holographic entanglement entropy proposal. In general time-dependent states, the proposal asserts that the entanglement entropy of a region  in the boundary field theory is given by a quarter of the area of a bulk extremal surface in Planck units.  The main element of our discussion is an implementation of an appropriate Schwinger-Keldysh contour to obtain the reduced density matrix  (and its powers) of a given region, as is relevant for the replica construction.  We map this contour into the bulk gravitational theory, and argue that the saddle point solutions of  these replica geometries lead to a consistent prescription for computing the field theory R\'enyi entropies. In the limiting case where the replica index is taken to unity, a local analysis suffices to show that these saddles lead to the extremal surfaces of interest. We also comment on various properties of holographic entanglement that follow from this construction.
}

\begin{document}

\maketitle

%~~~~~~~~~~~~~~~~~~~~~~~~~~~~~~~~~~~~~~~~~~~~~~~
\section{Introduction}
\label{sec:intro}
%~~~~~~~~~~~~~~~~~~~~~~~~~~~~~~~~~~~~~~~~~~~~~~

One of the intriguing aspects of the holographic AdS/CFT correspondence is the geometrization of quantum entanglement. In a local QFT the amount of correlation between degrees of freedom confined in a spatial region, and those outside, is measured by the entanglement entropy.  While simple to state, this quantity  is notoriously hard to compute in all but a handful of circumstances. At a technical level its computation requires determining the logarithm of a state-dependent operator (see below), which is  challenging  in an interacting QFT. 

Given this situation, it is rather remarkable that one has a rather simple way to compute entanglement in 
holographic field theories thanks to the AdS/CFT correspondence. Specifically, Ryu and Takayanagi (RT) \cite{Ryu:2006bv,Ryu:2006ef} argued, drawing analogy with the behaviour of black hole entropy in gravitational theories, that the entanglement entropy ascribable to a region $\regA$ is given by solving a classical geometric problem in AdS. One is instructed to find a minimal area surface anchored on the boundary $\entsurf$ of the region of interest; its area in Planck units measures the entanglement entropy $S_\regA$. The RT prescription was originally given for static states; this extends trivially to  more generally to  states at a moment of time reflection symmetry. However, the concept of entanglement being quite fundamental is not limited to such situations alone. Indeed the notion of entanglement entropy makes sense even when the state in question involves non-trivial temporal evolution. The Hubeny-Rangamani-Takayanagi (HRT) proposal \cite{Hubeny:2007xt} mitigates this lacuna by arguing that the correct extension of the RT prescription in time-dependent situations involves consideration of codimension-2 {\em extremal} surfaces. 

The primary intuition behind the HRT proposal was to ask what is the correct covariant generalization of the RT prescription. As we know from other aspects of gravitational physics, the principle of general covariance is a strong guiding principle, which in conjunction with other physical requirements serves to almost always zero in on the (oftentimes unique) dynamical construction. Demanding that the RT prescription admit a covariant upgrade, along with agreement in a common domain of applicability, led HRT to the extremal surface prescription.\footnote{ It should be noted that general covariance by itself is not strong enough \cite{Hubeny:2007xt}. For e.g., one can use causal structures to motivate a different construction leading to the  causal holographic information \cite{Hubeny:2012wa}. See also  \cite{Hubeny:2014qwa} for a discussion of the merits of general covariance as a guiding principle in a closely related context. }

Whilst these prescriptions for computing entanglement entropy in the holographic context are extremely simple,  one would like to derive them from first principles using nothing but the basic entries of the AdS/CFT dictionary. This has been achieved for the RT prescription explicitly by Lewkowycz and Maldacena (LM) \cite{Lewkowycz:2013nqa}  who mapped the replica construction usually employed to compute entanglement entropy in quantum field theories to the Euclidean quantum gravity path integral. In this context it is worth mentioning \cite{Casini:2011kv} who argued for the RT prescription in the case of spherical regions in the vacuum of a CFT  (see also \cite{Fursaev:2006ih} for an initial attempt at a proof). 

The argument forwarded by LM roughly proceeds as follows: given a density matrix $\rhoA$ a measure of entanglement is provided by the von Neumann entropy  $S_\regA = -\tr\left(\rhoA \, \log \rhoA\right)$. In practice, owing to the technical complexities of  taking the logarithm of an operator, one instead computes the R\'enyi entropies $S^{(q)}_\regA = \frac{1}{1-q} \, \log \tr (\rhoA^q)$ for $q \in {\mathbb Z}_+$ \cite{Renyi:1960fk}. Analytically continuing these R\'enyi entropies away from integral R\'enyi index $q$, one  obtains the von Neumann entropy in the limit $q \to 1$.  

When the density matrix is at a moment of time-reflection symmetry (or more simply just time-translationally invariant) one can employ Euclidean path integral techniques. One formally writes $\rhoA = e^{-2\pi \, K_\regA}$ which defines the modular Hamiltonian $K_\regA$ and views the computation of the  $q^{\rm th}$ R\'enyi
entropy as an evolution around an `Euclidean replica circle' parameterized by $\tau$ with $\tau \sim \tau + 2\pi\,q$.
This is entirely analogous to the computation of the canonical partition function by evolving along the Euclidean thermal circle for a period set by the inverse temperature before taking the trace. This analogy is in fact exact in the case of the spherical regions $\regA$ in the vacuum state of a CFT  \cite{Casini:2011kv},  for in this case the reduced density matrix  is unitarily equivalent to the thermal density matrix on hyperbolic space (by a conformal mapping).  
The LM construction can be viewed as the correct generalization of the global argument of \cite{Casini:2011kv} to situations where $\rhoA$ is not exactly thermal  (or equivalently the modular Hamiltonian is not local).

On the bulk side the R\'enyi entropies are computed by evaluating the on-shell gravitational action for the saddle point solution to the (Euclidean) quantum gravity path integral with asymptotic AdS boundary conditions set to include a replica circle of size $2\pi \,q$.\footnote{ As emphasized by LM \cite{Lewkowycz:2013nqa} the gravitational argument is quite general and transcends the specific application to deriving the RT prescription in the AdS/CFT context.}  If one however assumes that the  discrete ${\mathbb Z}_q$ replica symmetry which is respected in the field theory remains unbroken by the bulk saddle, then one can consider instead the gravitational action for the geometry obtained by a ${\mathbb Z}_q$ quotient.  This latter spacetime is singular; it has generically a codimension-2 conical singularity (with defect angle $\frac{2\pi}{q}$), as the replica circle is required to smoothly shrink in the original dual spacetime. The on-shell action is simply $q$ times that of the orbifolded geometry (with no contribution from the singular loci).

The main merit of this picture allows us to implement the analytic continuation away from $q\in {\mathbb Z}_+$ much more simply in the gravitational setting than in the field theory. We start with the geometry dual to the original density matrix $\rhoA$ and insert a conical defect of opening angle $\frac{2\pi}{q}$. Analyzing the local neighbourhood of the defect  for $q\to1^+$ one learns that a regular solution of gravitational equations of motion requires that the defect be a codimension-2 extremal surface, i.e, the trace of its extrinsic curvature in both normal directions vanishes.  In the replica $\tau$ direction this is a consequence of the time-reflection symmetry, but in the spatial normal direction, this statement is the origin of the minimal surface condition of RT. Once we have this basic statement, it then follows from diffeomorphism invariance that the variation of the gravitational path integral localizes on the defect (as $q\to 1$) and computes (in Einstein gravity) the area of this codimension-2 surface as desired.\footnote{ The RT (HRT) prescription also needs to be equipped with a statement that the minimal (extremal) surface is homologous to the boundary region $\regA$ \cite{Headrick:2007km} (see also \cite{Headrick:2013zda,Headrick:2014cta}). This does follow from LM provided that the extremal surface arises from a conical defect which for every $q \in {\mathbb Z}_+$ admits a lift to a smooth spacetime satisfying the boundary conditions of the replicated field theory \cite{Haehl:2014zoa}.  }

Let us now take stock of the discussion above. The essential ingredients used by LM can be distilled into: 
\begin{itemize}
\item   the entry in the AdS/CFT dictionary mapping field theory partition functions to a gravitational path integral with asymptotic boundary conditions \cite{Witten:1998qj}, 
\item  the assumption of unbroken replica symmetry, and 
\item  the use of Euclidean quantum gravity  techniques to evaluate the gravitational answer as the on-shell action (in a saddle point approximation).
\end{itemize}
The first two points are not quite specific  to situations with time-reflection symmetry and ought to apply more generally, provided we understand the third, i.e., formulate the computation of R\'enyi entropies in time-dependent situations carefully. 

To appreciate this point let us first ask what is involved in computing the matrix element of a time-dependent density matrix $\rhoA(t)$. Let us first start with the entire system and construct the density matrix (which may be pure) on $\regA \cup \regA^c$ at time $t$. Consider the pure case $\rho = \ket{\psi} \bra{\psi}$ for simplicity (if it is mixed, we can always purify it). Since a density matrix $\rho $ is an operator on the Hilbert space, we are required to evolve both the state vector and its dual from the initial state up to the time of interest $\rho(t) = {\cal U}_{t,-\infty}\, \rho_i \, {\cal U}_{-\infty, t}$, with ${\cal U}_{t_1,t_2}$ being the unitary operator that evolves the state forward from time $t_2$ to $t_1$, and $\rho_i$ being the initial state prepared in some manner at $t=-\infty$.\footnote{ Note that it is not essential that the initial state $\rho_i$ is prepared at $t=-\infty$; it could well be prepared at some finite $t_i$ (e.g.\ by a Euclidean path integral). However, in the following discussions we will set $t_i$ to $-\infty$ for linguistic simplicity.} As has been appreciated by many authors in the past, from a path integral perspective, one is necessarily led to doubling the degrees of freedom \cite{Schwinger:1960qe,Feynman:1963fq,Keldysh:1964ud}.  

% Figure 
\begin{figure}[htbp]
\begin{center}
\includegraphics[width=4in]{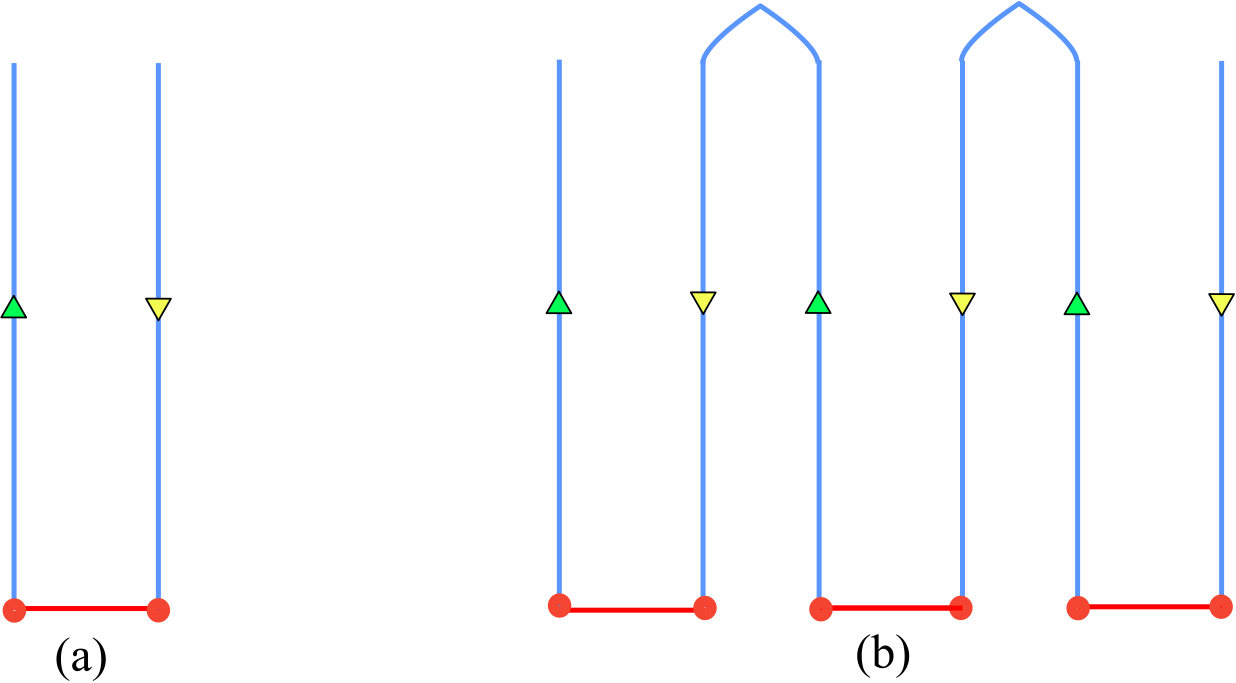}
\caption{A schematic representation of the Schwinger-Keldysh contours necessary for the computation of the (a) density matrix and (b) its powers. We have explicitly shown the computation of $\rho^3$ in (b). The dots and lines in red correspond to an entangled initial state prepared in some manner.  This picture does not carry the spatial information necessary to ascertain the reduced density matrices themselves, which is better understood from Figs.~\ref{fig:sk1}. Note that in contrast to the usual depiction of the Schwinger-Keldysh contour we will draw time running vertically.} 
\label{fig:sk0}
\end{center}
\end{figure}

So we start with two copies of the field theory and construct the instantaneous density matrix $\rho(t)$, see Fig.~\ref{fig:sk0}. Having done so we need to trace out the part of the system corresponding to $\regAc$. Since $\rhoA$ is a wedge observable \cite{Casini:2008wt} (see also \cite{Headrick:2014cta}), it cannot depend on which Cauchy slice we pick in the domain of dependence of the chosen region $D[\regA]$. In particular, we can think of tracing out $\regAc$ as setting boundary conditions on the past of $\regA$'s domain of dependence. Our choice of the Schwinger-Keldysh contour then allows us to immediately compute the matrix elements of $\rhoA$ -- see Fig.~\ref{fig:sk1} for a schematic representation of this construction (which will be explained in \S\ref{sec:qft}). Once we have the matrix element of $\rhoA$, we simply string together $q$ copies together cyclically. This is efficiently done in the path integral construction by prescribing an appropriate Schwinger-Keldysh contour, which as can be guessed will involve $2q$ copies of our field theory to account for both doubling and  replicating.

Once we have identified the field theory algorithm for the computation of the R\'enyi entropies, we can then ask what is the gravitational avatar of these Schwinger-Keldysh contours. This question has been addressed in the  AdS/CFT context a long time ago: a prescription to compute real time correlation functions was first given in  \cite{Son:2002sd} and was subsequently derived from a Schwinger-Keldysh framework in \cite{Herzog:2002pc}. More recently,  \cite{Skenderis:2008dh,Skenderis:2008dg}  gave a general prescription for computing real-time observables, focusing in particular on answering the question of interest to us: what is the gravity dual of a Schwinger-Keldysh contour? 

\begin{figure}[htbp]
\begin{center}
\includegraphics[width=5in]{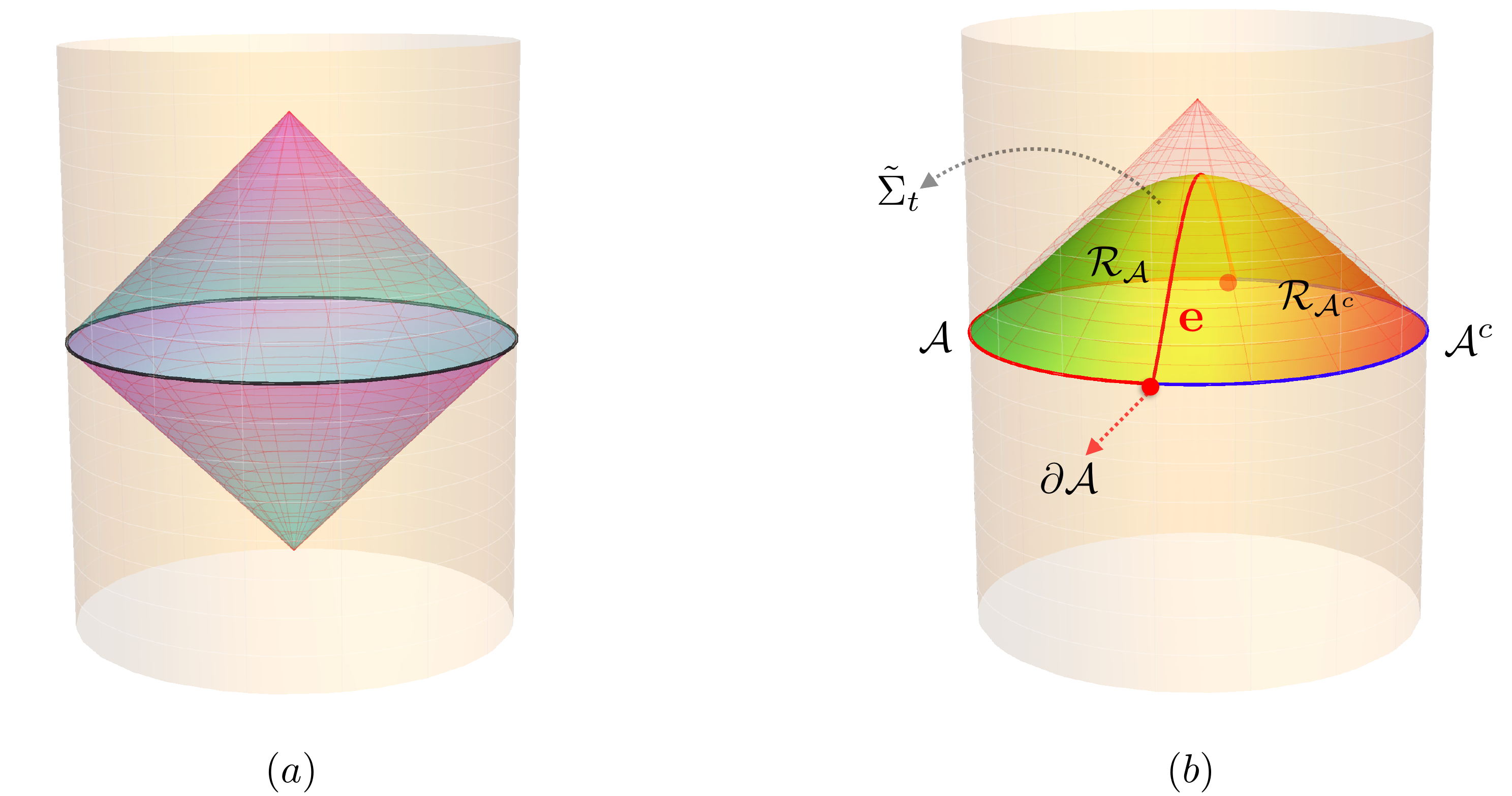}
\caption{The bulk domains of interest in the Lorentzian construction. (a) We show the Wheeler-DeWitt patch associated with a give Cauchy surface on the boundary. (b) Given a separation of the boundary Cauchy surface into regions $\regA$ and $\regAc$ respectively, any bulk Cauchy surface $\tilde{\Sigma}_t$ in the Wheeler-DeWitt patch admits a decomposition $\tilde{\Sigma}_t = \homsurfA \cup \homsurfAc$. We also display the bulk codimension-2 fixed point locus $\fixM$ anchored on the entangling surface which approaches the extremal surface $\extr$ as $q\to 1$.}
\label{fig:bulkDoms}
\end{center}
\end{figure}

To motivate their construction let us first realize one important fact about the AdS/CFT map.  For the boundary QFT, we are free to choose a background geometry which we take to be a globally hyperbolic Lorentzian manifold $\bdy$ for this purpose.  This ensures that we have a nice foliation of $\bdy$ by Cauchy slices $\Sigma_t$ and can thus compute observables at time $t$. A boundary slice $\Sigma_t$ however does not uniquely pick a corresponding time slice in the bulk. Instead, it may continue into one of infinitely many bulk Cauchy slices that all lie in the bulk region spacelike from $\Sigma_t$ itself, the  {\em Wheeler-DeWitt patch}
of $\Sigma_t$, cf., Fig.~\ref{fig:bulkDoms}. The gravity dual of the Schwinger-Keldysh contour \cite{Skenderis:2008dg} involves a bulk geometry which coincides with the original spacetime up unto one  of the bulk Cauchy slices in the Wheeler-DeWitt patch and then reverses its path onto a mirror copy of the bulk spacetime. Thus even the bulk spacetime is doubled in the process of giving  a gravitational construction of the real time contour, with the only proviso that there is an ambiguity in the choice of the bulk slice in the Wheeler-DeWitt patch where we reverse the trajectory. The ambiguity -- being unphysical -- must cancel out in physical observables; this was shown in \cite{Skenderis:2008dg} to be the case for computing correlation functions of local operators.

% Figure 

Since we are interested in working with a piece $\cal A$ of the boundary Cauchy slice, we have to generalize the above discussion to restrict attention to the past domain of dependence of $\regA$ on the boundary.
What this amounts to is that we start with the part of the spacetime which is relevant for computing $\rho(t)$ -- this involves slicing the geometry dual to $\ket{\psi}$ at some Cauchy slice within the Wheeler-DeWitt patch of $\Sigma_t$ and gluing back a second copy of the same. However, while arbitrary bulk Cauchy slices are acceptable for the computation of correlation functions in the state $\ket{\psi}$, for the computation of $\tr (\rho_{\cal A})^q$ we will need to restrict to special bulk Cauchy slices in the Wheeler-DeWitt patch of $\Sigma_t$. Specifically one requires that  the allowed bulk Cauchy slice contain the fixed points under the ${\mathbb Z}_q$ replica symmetry.

Once we realize this we are in a position to set up the fully Lorentzian construction of $\tr (\rhoA)^q$, involving as advertised $2q$ copies of the bulk spacetime glued together to reflect the boundary Schwinger-Keldysh construction, which computes the R\'enyi entropies in a way which is consistent with field theory causality . One can then invoke the replica symmetry to focus on a single unit of the Lorentzian forward-backward temporal evolution by taking the ${\mathbb Z}_q$ quotient of the resulting geometry. The question then reduces to ascertaining where the fixed point set of the ${\mathbb Z}_q$ replica symmetry, $\fixM$, lies and what ensures regularity in the $q\to 1$ limit. The locus $\fixM$  is the extension of the boundary fixed point set, viz., the entangling surface $\entsurf$, to the bulk and it has to be invariant with respect to unitaries both in $\regA$ and $\regAc$. Furthermore, since these geometries  correspond to the duals of the $q \rightarrow 1$ limit of $\tr (\rho_{\cal A})^q=\tr (\rho_{\regAc})^q$ and  $\fixM$ is common to both,  $\fixM$ cannot be in causal contact with $D[\regA]$ or $D[\regAc]$. If it were, that would mean that $\regA$ is in causal contact with $\regAc$ through the bulk \cite{Headrick:2014cta}. The upshot is  that $\fixM$ lies in the \emph{causal shadow} \cite{Headrick:2014cta}. The local analysis in the neighborhood of $\fixM$ ends up being essentially the same as the one used in the LM construction, except that the normal plane to the fixed point set has Lorentz signature. We argue that this leads to the extremal surface condition of HRT when the bulk dynamics is Einstein gravity  (or an appropriate generalization along the lines of \cite{Dong:2013qoa,Camps:2013zua} for theories of higher derivative gravity). Requiring that the boundary causality be respected by  the bulk dynamics, we end up with a physical result consistent with the discussion in  \cite{Headrick:2014cta}. 

The outline of this paper is as follows. We begin in \S\ref{sec:qft} where we flesh out the details of the Schwinger-Keldysh contour relevant for the computation of R\'enyi entropies in time-dependent (mixed) states of a quantum field theory.  We then move on in 
\S\ref{sec:grav} with the gravitational analogue of this field theory computation, reviewing in the process the LM construction in \cite{Lewkowycz:2013nqa} and the changes necessary for our argument. We demonstrate that the HRT proposal involving extremal surfaces follows from the bulk analysis, and sketch how the computation of the on-shell action in the Lorentzian context leads to the desired area functional. Arguing that we have a method for computing
all R\'enyi entropies -- at least in principle -- turns out to be a bit more subtle, but possible -- the technical steps necessary are given in Appendix \ref{sec:action}. In \S\ref{sec:discussion} we end with a discussion of how various aspects of the HRT proposal are manifested from our perspective.

%~~~~~~~~~~~~~~~~~~~~~~~~~~~~~~~~~~~~~~~~~~~~~~~
\section{Field theory construction}
\label{sec:qft}
%~~~~~~~~~~~~~~~~~~~~~~~~~~~~~~~~~~~~~~~~~~~~~~

We will begin our discussion by presenting a field theory construction of reduced density matrices in time-dependent states using the Schwinger-Keldysh (aka in-in) formalism \cite{Schwinger:1960qe,Keldysh:1964ud}. As explained in \S\ref{sec:intro} our motivation is to use this prescription to set up a path integral to compute R\'enyi entropies in non-trivial time-dependent states.

Without loss of generality let us consider a pure state of some $d$-dimensional local QFT on a globally hyperbolic background geometry $\bdy$.\footnote{ A mixed state can be purified by introducing appropriate auxiliary degrees of freedom.} We will prepare an initial state of this theory and evolve it up to some time $t$. The specifics of the initial state will not be important for our purposes; it could  be constructed either by invoking some boundary conditions in the far past or by slicing open a Euclidean path integral with possible sources. Regardless of how the state is prepared, we will evolve it in real (Lorentzian) time, further allowing ourselves the freedom to turn on arbitrary spacetime-dependent sources. 

%~~~~~~~~~~~~~~~~~~~~~~~~~~~~~~~~~~~~~~~~~~~~~~~
\subsection{The time-dependent QFT wavefunctional}
\label{sec:qftwf}
%~~~~~~~~~~~~~~~~~~~~~~~~~~~~~~~~~~~~~~~~~~~~~~

With this understanding let us attempt to write down a formal path integral expression for the wavefunctional of the QFT.
We first note that the density matrix requires information about both the state and its conjugate,  which have to be separately evolved from the initial state (perhaps with a time-dependent Hamiltonian in the presence of sources). Schematically and working in the Schrodinger picture, one has therefore
\begin{equation}
\begin{split}
\ket{\Psi(t)} = e^{i\,H \,t} \ket{\Psi_0}\,, \qquad \bra{\Psi(t)} = \bra{\Psi_0}  e^{-i\,H \,t}   \\
\qquad \Longrightarrow \qquad \rho(t) = \ket{\Psi(t)} \bra{\Psi(t)} = e^{i\,H \,t} \, \rho_0 \,e^{-i\,H \,t} \,.
\end{split}
\label{eq:rhoev}
\end{equation}

It is useful for the purpose of our discussion to convert this statement into a path integral construction. A natural way to respect 
\eqref{eq:rhoev} is to incorporate a Keldysh contour \cite{Keldysh:1964ud} that evolves the state $\ket{\Psi} $ forward and its conjugate $\bra{\Psi}$ backward. In the path integral construction, the wavefunctionals are given as
\begin{equation}
\begin{split}
\Psi(t,\phi_0(x)) &=\int [D\phi]\,  e^{i S[\phi]_{-\infty}^t} \, \delta(\phi(t,x)-\phi_0(x)) \,,\\ 
\bar{\Psi}(t,\phi_0(x)) &=\int [D\phi] \, e^{i S[\phi]^{-\infty}_t} \, \delta(\phi(t,x)-\phi_0(x)) \,,\end{split}
\label{eq:wavfns}
\end{equation} 
where $\phi$ stands for the entire collection of fields of the theory, and $S[\phi]_{t_1}^{t_2}$ is defined as $\int_{t_1}^{t_2} dt L[\phi]$ for the QFT Lagrangian $L$. It is then easy to write down the corresponding path integral expression for $\rho(t)$ as:
\begin{equation}
\rho = \ket{\Psi}\bra{\Psi} = \int [D\phi] \, e^{i\,S_\uparrow[\phi] - i\, S_\downarrow[\phi]}
\label{eq:rhopi}
\end{equation}  
where the arrows indicate the direction of time evolution inherited from \eqref{eq:rhoev}. As is well appreciated in the literature, this can be viewed quite straightforwardly in terms of a contour prescription for the path integral, cf., Fig.~\ref{fig:sk1} for an illustration.\footnote{ To get physical answers we usually have to implement an appropriate $i\varepsilon$ prescription in case we encounter singularities along the real time axis  (which is the naive contour of integration); this is implicit in the Schwinger-Keldysh contour.}

%~~~~~~~~~~~~~~~~~~~~~~~~~~~~~~~~~~~~~~~~~~~~~~~
\subsection{The reduced density matrix} 
\label{sec:rhoA}
%~~~~~~~~~~~~~~~~~~~~~~~~~~~~~~~~~~~~~~~~~~~~~~

Having constructed the path integral representation for the total state of the system $\rho(t)$ at the instant of our choosing, we now want to use this to compute the entanglement entropy for a subsystem at this instant.  Let us denote the achronal spatial section (a Cauchy slice) of $\bdy$ at this instant as $\CSA$. The subsystem of interest will be a spatial region $\regA \subset \CSA$ bounded by the {\em entangling surface} $\entsurf$. We denote the complementary region as $\regAc$; clearly $\Sigma_t = \regA \cup \regAc$. Of interest to us is the reduced density matrix $\rhoA(t)$ induced on $\regA$ from the global state $\rho(t)$ on $\CSA$; subsequently we will want to compute various trace class observables of $\rhoA$.\footnote{ By trace class observables, we mean expectation values of operators localized in $\regA$ (more precisely in its past domain of dependence $D^-[\regA]$, cf., footnote \ref{fn:causal}), i.e., $\tr\left(\rho_\regA\, {\cal O}_\regA\right)$ where ${\cal O}_\regA$ is a collection of operators with support in $\regA$. } 

\begin{figure}[h!]
\begin{center}
\vspace{5mm}
\includegraphics[scale=0.6]{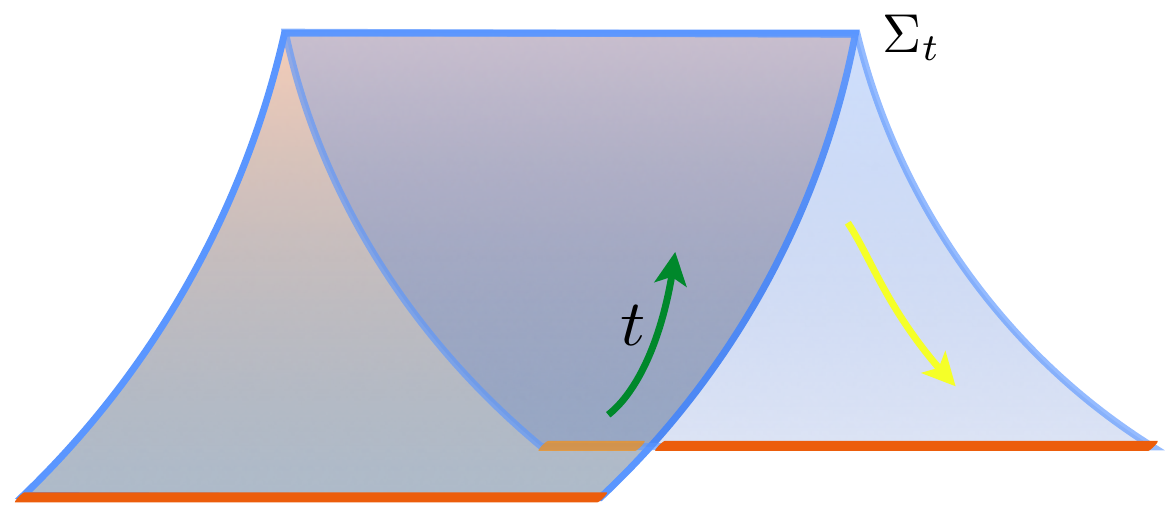}
\hspace{5mm}
\includegraphics[scale=0.6]{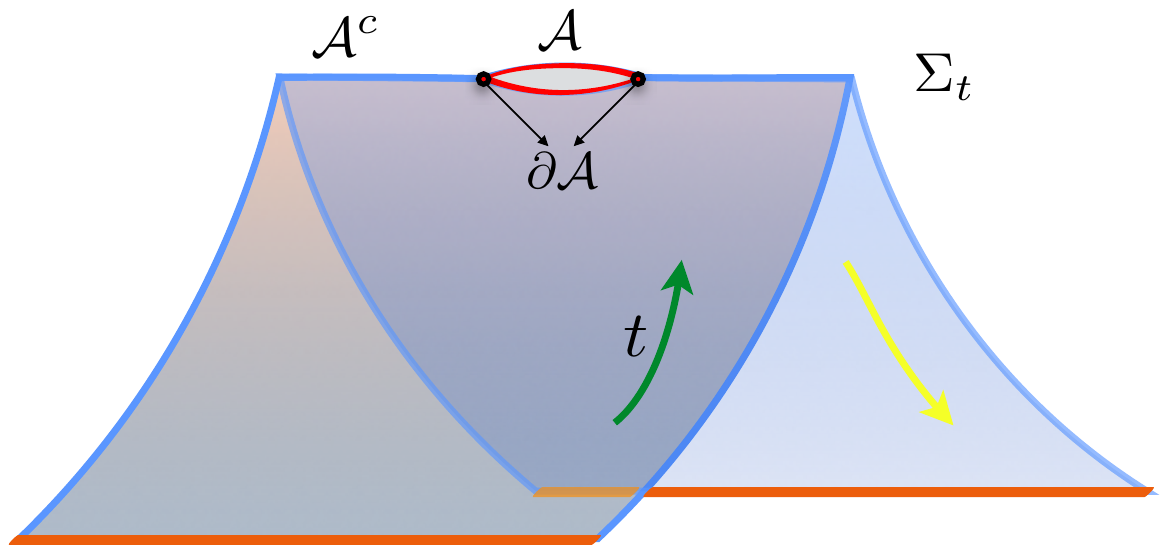}
\vspace{5mm}
\caption{Schwinger-Keldysh construction for $\tr \rho(t)$ and $\rhoA$. The forward evolution for $\ket{\Psi}$ proceeds up to 
$\CSA$, while the backwards evolution for  $\bra{\Psi}$ starts there. Gluing the two evolutions together at $\CSA$ enables taking the trace of $\rho(t)$. We also depict the situation, where we open out this trace along $\regA \subset \CSA$ to construct $\rhoA(t)$. These cuts are introduced at $t= t^\pm$ as described in the main text.}
\label{fig:sk1}
\end{center}
\end{figure}

The bipartition of $\CSA$ allows the computation of the matrix elements of $\rhoA$. Recall that $\rho(t)$  has been obtained by combining two separate evolutions, one forward and one backward. Gluing these two  evolutions together along the Cauchy surface $\CSA$ ends up computing $\tr(\rho(t))$. At this point it is useful to upgrade the conventional pictures of Schwinger-Keldysh contours to depict spatial information as well; see Fig.~\ref{fig:sk1}. We now mark off $\regA$ on $\CSA$ and introduce some additional boundary conditions to extract the reduced density matrix $\rhoA$.

To obtain the matrix elements of $\rhoA(t)$, we introduce a cut along $\regA$ in the path integral by integrating over $\phi_0$ only in region $\regAc$; the cut however remains pinned at $\entsurf$.  We imagine that the cut is introduced at $\CSA$, which we will denote $t = t^\pm$ for the two legs of the evolution respectively. Since there are two temporal contours in the path integral we have two cuts, which open up a slit in the folded geometry as depicted in Fig.~\ref{fig:sk1}.
On either opening of the slit, we prescribe some boundary conditions for the fields $\phi$ along $\regA$: $\phi(t =t^+)\big|_{\regA} = \phi_+$ in the forward part of the contour and $\phi(t =t^-)\big|_{\regA} = \phi_-$ in the backward part. The path integral with these boundary conditions for the fields on the cuts across $\regA$ defines for us the matrix element  $(\rhoA)_{+-}$.

While our choice of boundary conditions to construct the reduced density matrix disconnects the path integral contour across $\regA$, it remains geometrically connected across the complementary region as we go through $\CSA$. If we want to compute $\tr(\rhoA)$, we can simply glue back the slit that we introduced and therefore recover the original (doubled) geometry. It will be important in what follows that the local geometry in the neighbourhood of the entangling surface $\entsurf$ is flat; in the Lorentzian context of interest we could say it is ``locally Rindler-like''.

The construction of the matrix element of $\rhoA$ involves picking a Cauchy slice. However, we have some freedom here; the precise choice of $\CSA$ is relevant only for the computation of the density matrix itself. 
For computing trace class observables, we can use unitary evolutions within causal domains of relativistic QFTs (see e.g., \cite{Headrick:2014cta}) to pick other slices.\footnote{ We use standard relativistic notations: $J^\pm[p]$ stands for the causal future/past of $p$ and $D^\pm[X]$ the future/past domain of dependence of some set $X$.  \label{fn:causal}}
Computation of traces involving $\rhoA$ only requires information about the temporal evolution of the state in the past of the entangling surface. Thus we can push parts of the Cauchy slice (and thus $\regA$, $\regAc$) to the past as long as we do not modify the entangling surface. More formally, replacing $\Sigma_t$ by another $\Sigma'_t \subset J^-[\Sigma_t]$ with $\Sigma_t' \cap \Sigma_t = \entsurf$ is acceptable.\footnote{ One can use this observation to deform $\CSA$ down all the way to a spacelike surface that lies arbitrarily close to the null surface ${\cal N}_D =\partial D^{-}[\regA] \cup \partial D^{-}[\regAc]$ and replace the wave functional on $\Sigma_t$, $\Psi(\phi_{\pm})$, by the wave functional on ${\cal N}_D$, $\Psi(\phi^D_{\pm})$, obtained by Hamiltonian evolution. This perspective is useful for some purposes.}

Before we proceed to construct the geometries relevant for computing the R\'enyi entropies, let us introduce some coordinates to describe the folded geometry for $\rho_{\regA}$. A useful coordinate chart is the one inspired  by the fact that the local geometry in the vicinity of $\entsurf$ is Rindler space (as opposed to a Cartesian chart around $\CSA$). To see this consider the following example: in flat spacetime $\bdy = {\mathbb R}^{d-1,1}$ we choose the Cartesian coordinates $(t,x,w^i)$, $i=1,2,\cdots, d-2$. Take $\CSA$ to be a constant time slice (say $t=0$) and $\regA$ to be the half space defined by positive $x$. In this case $D^-[\regA]$ is the past half of the (right) Rindler wedge of flat space. The construction described above can be implemented in this case.\footnote{ In this highly symmetric example, one usually recognizes that it has a timelike Killing field and exploits the time-reflection symmetry. We phrase the statements herein without invoking this symmetry, therefore enabling generalization to the time-dependent situation of interest. } 

\begin{figure}[h!]
\begin{center}
\vspace{5mm}
\includegraphics[scale=0.7]{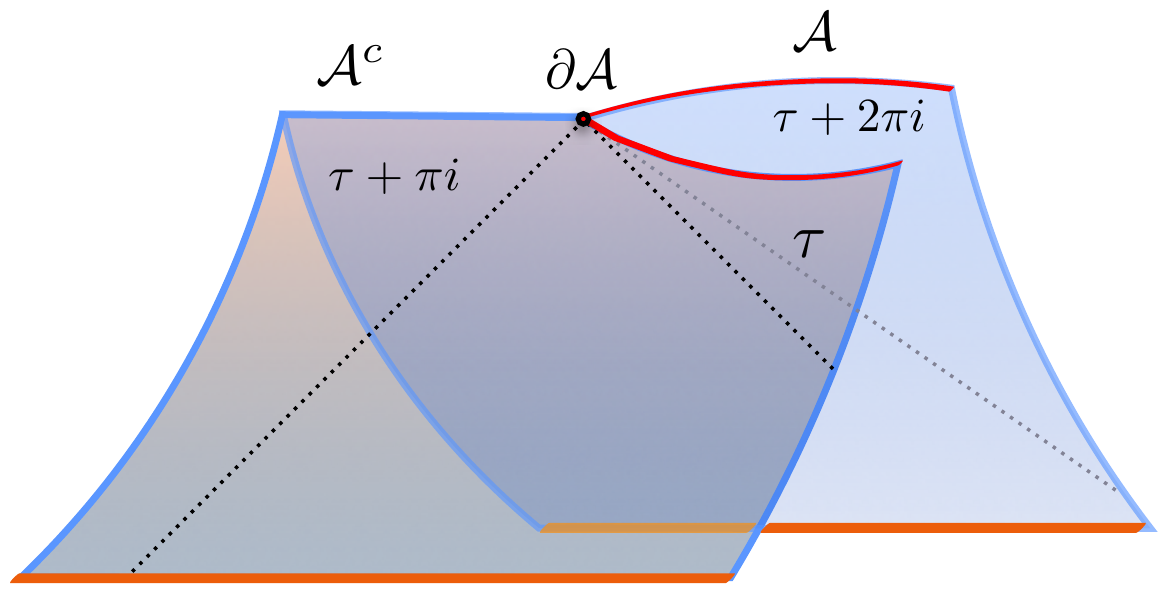}
\vspace{5mm}
\caption{Local Rindler coordinates that we will use to describe the geometric construction of 
$\rho_{\regA}$. We have focused on the neighbourhood of the entangling surface and indicated the causal domains and coordinates used therein (see text for details). }
\label{fig:rindler}
\end{center}
\end{figure}

Let us therefore consider the Rindler chart:  
\begin{align}
ds^2 =-dt^2 + dx^2 +  dw^i\, dw^i   = - r^2\, d\tau^2 +dr^2+  dw^i\, dw^i \,.
\label{eq:rindler1}
\end{align}
The advantage of these Rindler coordinates is that we can simultaneously refer to all the spacetime regions in question, by allowing $\tau$ to be complex with a discrete imaginary part\footnote{ It is worth emphasizing that this is simply a convenient book-keeping notation. In particular, we are not assuming any analytic properties on the complex $\tau$ plane.}.  Since we are dealing with the reduced density matrix, for $x>0$, $t^{\pm}$ are disconnected. Consider then the five coordinate patches covered by  $\tau = \tauA +\frac{m\,\pi}{2}\, i$, with $\tauA$ real and $m=0,1,2,3,4$ (see Fig.~\ref{fig:rindler} for an illustration):\footnote{ We need five patches because we are describing the reduced density matrix in its entirety. When taking trace we will end up identifying the patches under $\tau \to \tau+2\pi i$.} 
\begin{itemize}
\item $m=0$: We use $\tau =\tauA<0 $ to coordinatize the domain $D^-[\regA]$ below the fold at the Cauchy slice $\CSA$.  One may think of these domain as the right Rindler wedge, with boundary conditions  $\phi_{+}$ at 
$\regA$.
% $\Sigma_{\regA}$,. 
\item $m=1$: $\tau = \tauA + \frac{\pi}{2} i$ coordinatizes the Milne wedge $J^-[\entsurf]$  on the backwards segment of the Schwinger-Keldysh contour.
\item $m=2$: $\tau = \tauA+\pi i$ covers the past domain of dependence of the complementary region $\regAc$. Since we are describing $\rho_{\regA}$, the two folds are glued at $\CSA$.  The forward and backward domains of $D^-[\regAc]$ are distinguished because $\tauA$ is positive/negative in the backwards/forwards segment \footnote{ Note that the signs are flipped with respect to the right wedge, as it is usual in Rindler.}. 
\item $m=3$: When $\tau = \tauA+ i\frac{3\pi}{2}$  we are in the Milne wedge again $J^-[\entsurf]$, but this time on the forward segment of the Schwinger-Keldysh contour.
\item $m=4$: This corresponds to  $\tauA >0 $ and coodinatizes the region $D^-[\regA]$ in the forward part of the contour. This is the right Rindler wedge again, but on the backward part of the Schwinger-Keldysh contour and we impose boundary conditions $\phi_-$ at $\regA$. 
\end{itemize}

It will be crucial to remember that when we  compute  $\tr(\rhoA)$ we glue the $m=0$ domain with the $m=4$ domain along $\regA$.
What this amounts to is a prescription for the identification $\tauA^{0^-} =\tauA^{0^+}$ in the path integral computing the trace of the reduced density matrix, since we want to think of the backward/forward parts of the right wedge as being parametrized by a unique coordinate.  We identify $
\tau \sim \tau +2\pi i $. This is the geometric encoding of the statement of $\entsurf$ being locally flat. One can equivalently phrase this by invoking the standard Rindler interpretation of a local accelerated observer seeing a temperature $\frac{1}{2\pi}$; cf., Fig.~\ref{fig:rindler}. 

 The local geometry near $\entsurf$ for more complicated geometries is similar; all that we require is that the normal bundle to  $\entsurf$ admits Lorentzian sections. In an open neighbourhood of the zero section i.e., at $\entsurf$ we can use the Rindler coordinates described above, which provides a convenient way to keep track of the properties of the replicated geometry.

%~~~~~~~~~~~~~~~~~~~~~~~~~~~~~~~~~~~~~~~~~~~~~~~
\subsection{Products of reduced density matrix and R\'enyi entropies}
\label{sec:}
%~~~~~~~~~~~~~~~~~~~~~~~~~~~~~~~~~~~~~~~~~~~~~~

At the end of the day, the reduced density matrix is a means for us to compute the entanglement entropy via the replica trick. We recall the basic definitions: 
\begin{equation}
\begin{split}
S_\regA &= - \tr_\regA\left(\rhoA \log \rhoA\right)  = \lim_{q \to 1} S_\regA^{(q)} \,,\\
S^{(q)}_\regA &= \frac{1}{1-q} \log \tr_\regA( \rhoA)^q \,.
\end{split}
\label{eq:vnren}
\end{equation}  
We have introduced the R\'enyi entropies, which help in implementing the replica trick via the second equality in the first line.

\begin{figure}[h!]
\begin{center}
\vspace{5mm}
\includegraphics[scale=0.45]{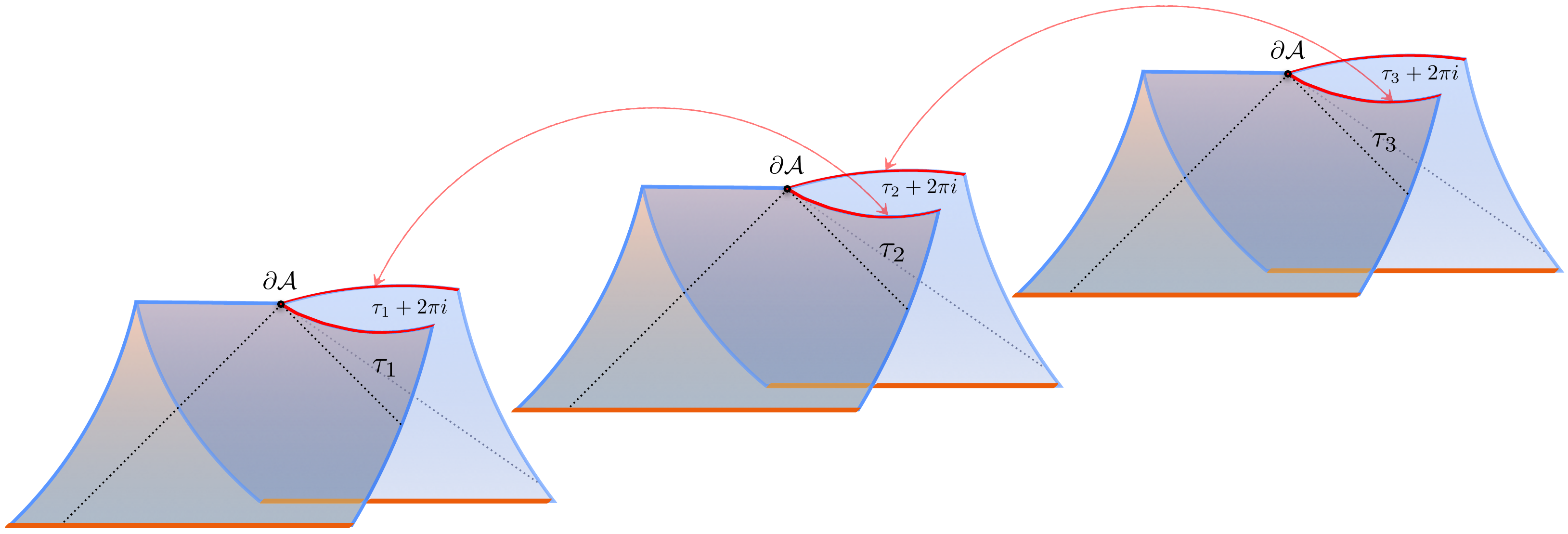}
\vspace{5mm}
\caption{A schematic view of the replica Schwinger-Keldysh contour for computing $(\rhoA)^3$  matrix elements. We have restricted attention to the neighourhood of the entangling surface for simplicity.  }
\label{fig:sk3renyi}
\end{center}
\end{figure}

To compute the  entanglement entropy we therefore need to compute powers of the reduced density matrix 
$(\rhoA)^q$. Since we have a path integral construction of $\rhoA$ this is easily done. Consider $q$ copies of the geometry used to compute the reduced density matrix elements $(\rhoA)_{+-}$ indexed by $I = 1, \cdots, q$.
Computing the matrix product requires us to cyclically glue these geometries together; this constructs the new field theory background $\bdy_q$. The construction involves identifications along the cuts introduced at $\regA$: we identify $\phi_{I-} = \phi_{(I +1)+}$. This should be interpreted as providing not only the boundary conditions for the path integral, but also giving us a geometric way to obtain a Schwinger-Keldysh contour which is relevant for the computation of $(\rhoA)^q$. In short one may view the thus obtained path integral contour as the Lorentzian analogue of the replica construction, see Fig.~\ref{fig:sk3renyi}. The latter of course  was useful in deriving the RT proposal from a gravitational path integral \cite{Lewkowycz:2013nqa}. Our task will be to abstract sufficient lessons from the Lorentzian construction in field theory to implement the same in gravity.

It is worth noting in passing that in certain situations where we have a time reversal symmetry $t\to-t$, we may simply evolve from $t=-\infty$ to $t=\infty$ on each copy of the replica construction.  We may then analytically continue to Euclidean signature on each copy. This is really the realm where the RT construction \cite{Ryu:2006bv} and its Euclidean quantum gravity derivation a la LM \cite{Lewkowycz:2013nqa} is valid. While we will review this in \S\ref{sec:lmreview}, our goal is to unanchor ourselves from this special circumstance and derive the covariant proposal of \cite{Hubeny:2007xt}.

Let us understand the geometry of interest, $\bdy_q$.  In the gluing construction, we have initially $2q$ different time coordinates (two for each copy of the density matrix, and $q$  density matrices), which are glued together differently in $\regA$ and $\regAc$ respectively. Instead of using different time coordinates, $t_I$, for each $I\in \{1,\cdots, q\}$, it is useful to invoke the $\tau$ coordinate inspired by the Rindler construction. Passing from $t_I \to \tau_I$ in each copy of the density matrix as described around \eqref{eq:rindler1}, the gluing involves identifying 
$\tau_J = \tau_{J-1} + 2\pi i$ along $\regA$. We now introduce a single coordinate $\tau$ with an imaginary part supplied to take care of this identification. Consider $\tau=\tau_{\regA} +  \frac{m \pi}{2}\, i$ where now $m =0, 1\, \cdots , 4q$. This coordinate naturally allows for interpolation from $\tau_1 \to \tau_2 \to \cdots \tau_q \to \tau_1$; the last identification ensures that we take the trace. What this implies is that $\tau \sim \tau + 2\pi i\, q$ is the correct identification for this coordinate along $\regA$ to ensure that the density matrices are multiplied in the right cyclic order. A pictorial depiction of this statement is presented in Fig.~\ref{fig:sk3renyi}. This statement is the Lorentzian analogue of the monodromy acquired by traversing sections of the normal bundle across $\entsurf$.

While a priori the choice of  coordinates is just a matter of convenience, the $\tau$ coordinates allow for a simple statement of the boundary conditions in the QFT. They should be seen as a useful book-keeping device for the identifications between the different replicas.  In particular, one is not performing any analytic continuation of the field theory data to complex times, as can be inferred by working directly with the Minkowski chart
using  $\{x_J, t_J\}$ for each copy of the reduced density matrix. 

The periodicity in $\tau \sim \tau + 2\pi i\,q$ is a statement about the gluing conditions in the Schwinger-Keldysh path integral contour constructed above. The field theory path integral is done over the $4q$ temporal domains, which are conveniently encapsulated by the single $\tau$ coordinate. We would like to reiterate what it means for fields to be periodic in $\tau$ with period $2\pi i q$.  In the QFT path integral we integrate over all allowed field configurations.  The $\tau$ domain of the fields consists of $4q$ disconnected horizontal lines in the complex $\tau$ plane, each with imaginary part $i\frac{\pi}{2}m, m=0, 1, \cdots, 4q-1$.  We then impose boundary conditions at the asymptotic infinities of these horizontal lines.  It is easiest to do this in the language of asymptotically incoming/outgoing modes.  The coefficients of these mode must match between $\tau=m \pi i+\infty$ and $\tau=(m+\frac{1}{2}) \pi i+\infty$, and between $\tau=(m+\frac{1}{2}) \pi i-\infty$ and $\tau=(m+1) \pi i-\infty$, for all $m=0,1,\cdots,2q-1$.  Note that in saying this, we are identifying $\tau = 2\pi i q-\infty$ with $\tau=-\infty$ -- this is what we mean by the periodicity $\tau \sim \tau+2\pi i q$. To recap: we are gluing $q$ copies of the reduced density matrix, with region $\regA$ identified across the copies. This can be just as well stated in the  $\{x_J, t_J\}$ coordinates, but the $\tau$ coordinate is more useful for delineating the analogous boundary conditions in gravity.

While the general focus here is on the computation of the R\'enyi entropies themselves, it will transpire that the gravitational computation is nicer for the derivative of the R\'enyi entropy with respect to its index. Define thus a related quantity 
\cite{Dong:2016fnf}, which we will call the \emph{modular entropy}:
\begin{equation}
\tilde{S}_\regA^{(q)} = -q^2\, \partial_q\left[ \frac{1}{q}\, \log \tr_\regA( \rhoA)^q\right]
\label{eq:renyiqder}
\end{equation}  
In writing this expression we have already assumed that we can analytically continue the R\'enyi entropies away from the integer values of the index  $q$. To our knowledge, this object has not been considered before in the quantum information literature, but it is rather natural. For instance, if we take $\rhoA$ to be of thermal  (as for spherical domains for CFTs in ${\mathbb R}^{d-1,1}$), and view $q$ to be a measure of the inverse temperature, then $\tilde{S}_\regA^{(q)} $ is the thermal entropy. The $q^{\rm th}$ modular entropy is the the appropriate Legendre transform of the $q^{\rm th}$ R\'enyi entropy.

Finally,  there is a ${\mathbb Z}_q$ symmetry relating the various replicas, with  $\partial \regA$ being the fixed point of its geometric action on $\bdy_q$.  In the rest of the paper, we will assume that this symmetry is unbroken.  As a consequence the one-point functions of our QFT should be replica symmetric. In the $\tau$ coordinate this corresponds to functions being strictly periodic with a smaller period of $2\pi i$ . As we will see later, this point of view provides a  particularly straightforward route to understanding the boundary conditions for the dual gravitational problem. At the end of the day we will require all fields in the bulk to be invariant with respect to the replica symmetry. Coupled with the fact that we disallow any curvature singularities, this serves to pick out the acceptable geometries, which satisfy Einstein's equation, and provide the dominant contribution to the gravitational action.

%~~~~~~~~~~~~~~~~~~~~~~~~~~~~~~~~~~~~~~~~~~~~~~~
\section{Gravitational construction}
\label{sec:grav}
%~~~~~~~~~~~~~~~~~~~~~~~~~~~~~~~~~~~~~~~~~~~~~~

We have thus far set up the problem of determining the matrix elements of the reduced density matrix and its powers by invoking an appropriate Lorentzian Schwinger-Keldysh contour prescription in the field theory. Assuming that the field theory in question is holographic, with a semi-classical gravitational dual described by a diffeomorphism invariant local Lagrangian, we would now like to ask how to implement the aforementioned computation in gravity. We will outline the construction below by first asking what is the gravity dual of the Schwinger-Keldysh contour of interest. This question was answered in \cite{Skenderis:2008dh,Skenderis:2008dg} whose analysis will inspire us to provide a bulk prescription for the computation of the reduced density matrix elements. Following this, assuming unbroken replica symmetry, we will argue that the entanglement entropy of the said density matrix is computed in terms of the area functional on a codimension-2 extremal surface in the geometry, thus deriving the HRT prescription \cite{Hubeny:2007xt}.

To set the stage for our discussion, we will first quickly review the elements that enter the derivation of generalized gravitational entropy \cite{Lewkowycz:2013nqa}, valid for ${\mathbb Z}_2$ time-reflection symmetric states.

%~~~~~~~~~~~~~~~~~~~~~~~~~~~~~~~~~~~~~~~~~~~~~~~
\subsection{Review: Time reflection symmetric case}
\label{sec:lmreview}
%~~~~~~~~~~~~~~~~~~~~~~~~~~~~~~~~~~~~~~~~~~~~~~

As explained in \S\ref{sec:qft}, if we have a $\mathbb{Z}_2$ time-reflection symmetry $t \rightarrow -t$ 
and are interested in computing the density matrix elements at the fixed point $t=0$, then the Schwinger-Keldysh path integral can be simplified. Exploiting the symmetry, we can simply consider the evolution from 
 $t=-\infty$ to $t=\infty$, since the backward evolution is equivalent to the forward one. This in turn  allows us
 to simplify the computation by analytic continuing to Euclidean time $t \to i\, \tE$.  

 Once we have a Euclidean path integral for computing matrix elements of $\rhoA$, the process of computing $\tr (\rhoA)^q$ is achieved by gluing the replicas in Euclidean space. Equivalently, one is instructed to compute the partition function of the field theory in a geometry $\bdy_q$ with conical excess   $2\pi\,q$ inserted at $\entsurf$. This turns out to be a well-defined gravitational problem. All one needs is to construct a bulk geometry $\bulk_q$ whose conformal boundary is $\bdy_q$ for $q\in {\mathbb Z}_+$. While this serves to compute the R\'enyi entropies, in fact, we are interested in computing the entanglement entropy, which is achieved by an analysis in the limit $q\to 1^+$. The key point of \cite{Lewkowycz:2013nqa} is that the analytic continuation from integral $q$ to the vicinity of $q \sim 1$ is much simpler in the gravitational context. We now review this argument, splitting it into two convenient parts: a purely kinematic piece and one that cares about the gravitational dynamics.

\paragraph{Kinematics:} Let us first discuss the case $q \in {\mathbb Z}_+$. For integer $q$, the boundary manifold $\bdy_q$ is a $q$-fold branched cover over $\bdy$ (branched at $\entsurf$). Per se this provides a clean boundary condition for the gravity problem as described above. However, we can exploit that fact that the partition function has 
%there is 
a ${\mathbb Z}_q$ symmetry of $\bdy_q$ that exchanges the different replicas. This is a symmetry owing to the cyclicity of the trace in the definition of R\'enyi entropies.  

Assuming as in \cite{Lewkowycz:2013nqa} that this  replica ${\mathbb Z}_q$ symmetry extends to the bulk, we can take the smooth bulk dual $\bulk_q$ and consider the quotient space ${\hat \bulk}_q = \bulk_q/{\mathbb Z}_q$. This quotient geometry is not smooth and generically contains a codimension-2 fixed point locus   of the ${\mathbb Z}_q$ action.\footnote{ There are some subtleties with this statement, for it is possible in certain situations that the fixed point set has `wrong' codimension; cf., \cite{Haehl:2014zoa} for a detailed discussion and examples. We will assume that we have a family of replica symmetric geometries, parameterized by $q$, and smooth for $q\in {\mathbb Z}_+$ which, as argued there, is sufficient to avoid any exotic scenarios.} We will call this fixed point set of the bulk $\fixM_q$ -- it will be part of the kinematic data as we build up an ansatz for construction. Apart from being invariant under the ${\mathbb Z}_q$ symmetry exchanging the replicas, $\fixM_q$ is the natural extension of $\entsurf$ into the bulk. 
\begin{figure}[h!]
\begin{center}
\vspace{5mm}
\includegraphics[scale=0.5]{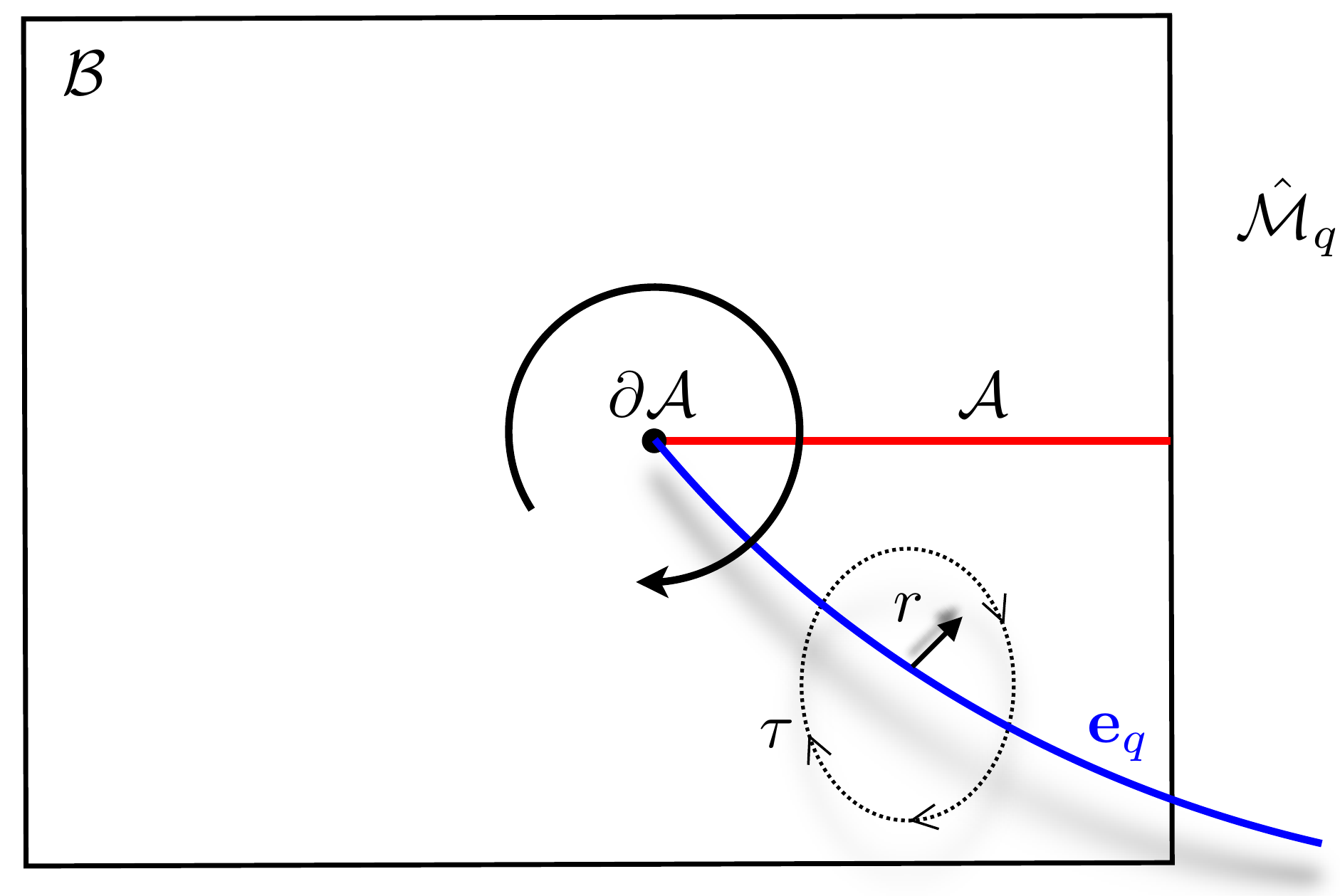}
\vspace{5mm}
\caption{Illustration of the local geometry near the fixed point locus of the replica ${\mathbb Z}_q$ symmetry action on the boundary and the bulk. The region $\regA$ terminates on the entangling surface $\entsurf$, which extends in ${\hat \bulk}_q$ into a fixed point locus $\fixM_q$. We use polar coordinates $(r,\tau)$ to parametrize sections of the (Euclidean) normal bundle of this fixed point set. }
\label{fig:bdycondq}
\end{center}
\end{figure}

Let us now set up a bulk coordinate chart. First, consider a codimension-2 surface in the original spacetime $\bulk$. We pick coordinates adapted to the surface: $y_i$ with $i=1,2,\cdots, d-1$ parameterize tangential directions, while the normal directions are coordinatized by $x, \tE$. Expanding the metric in a derivative expansion around the surface, we have
\begin{equation}
ds_{_ E}^2=dx^2+d\tE^2+\left(\gamma_{i j}+2\, K_{ i j}^x\, x+2\, K_{ i j}^t\, \tE \right)dy^i\, dy^j+ \cdots \,.
\label{eq:lm0}
\end{equation}  
We have retained only the leading terms in the Taylor expansion about the surface located at $x=0, \tE=0$. One can equivalently parameterize the normal directions in polar coordinates $x \pm i\,\tE= r \,e^{\pm i \,\tau}$, where $\tau \sim \tau +2\pi$ for regularity.\footnote{ For convenience we are going to use the same notation for the normal bundle coordinates in the bulk $\bulk$ and the boundary $\bdy$. This is natural; as the fixed point set $\fixM$ is the bulk extension of the entangling surface.}

If we introduce such local coordinates in the vicinity of $\fixM_q$ as in \eqref{eq:lm0}, then the replica symmetry implies that the action is invariant with respect to a global shift of the polar coordinate in the normal plane $\tau$, viz., $\tau \to \tau + 2\pi$; see Fig.~\ref{fig:bdycondq} for an illustration.  Near $\fixM_q$ this replica coordinate has to be identified under $\tau \sim \tau + 2\pi\,q$. We can now use the smoothness of the covering space $\bulk_q$,\footnote{ One might worry that the geometry is smooth in the bulk, but  becomes singular as it approaches the boundary due to the entangling surface. This singularity however can be dealt with by a suitable regularization procedure. For example in some situations \cite{Casini:2011kv} we can use conformal mapping to send $\entsurf$ to infinity and use a standard IR cut-off.} to infer that the local geometry near $\fixM_q$ in the quotient ${\hat \bulk}_q$ has be of the form\footnote{ Strictly speaking the geometry has a fibration structure, whereby the normal bundle parameterized by the $(r,\tau)$ coordinates is non-trivially fibred over the base. We have for simplicity dropped some of the off-diagonal components in writing \eqref{eq:lm1}.}
\begin{equation}
ds^2=q^2 \,dr^2+r^2 \,d\tau^2+ ds^2_\text{transverse}+...
\label{eq:lm1}
\end{equation}  
We have left implicit here the transverse part of the geometry which we will describe in due course. The main point to note is  the explicit $q$ dependence. Its presence implies that in order for  the metric to be smooth near $r=0$, we must encounter some non-trivial backreaction;  one cannot simply identify  $\tau \sim \tau+2\pi q$ in \eqref{eq:lm0}.

Exploiting the replica symmetry we can restrict our attention to a single fundamental domain (or replica) of the ${\mathbb Z}_q$ action in the bulk. Thence, the total action of the gravity computation will be $q$ times that of a single domain, viz.,
\begin{align}
I[\bulk_q] = q\, I[{\hat \bulk}_q]
\label{eq:lmI1}
\end{align}
While the quotient space has a conical singularity with defect angle $\frac{2\pi}{q}$, the covering space is smooth; this observation will play a crucial  role in setting up the boundary conditions.

The advantage of thinking about the orbifolded quotient space becomes manifest when we think about computing the entanglement entropy which requires analytic continuation from $q \in {\mathbb Z}_+$ to $q =1$. From the geometric perspective $q$ is simply a parameter that tells us the strength of the opening angle at the conical defect in ${\hat \bulk}_q$. Working in the orbifolded space, we simply analytically continue $q$ by dialing the strength of the singularity. This the kinematic part of the analysis implies that we work in the $q\to 1$ limit, on a geometry with a conical deficit of prescribed strength, with the same boundary conditions as the original background geometry $\bulk$.

\paragraph{Dynamics:} Having set up the basic problem in the gravitational context, we now want to figure out what configurations dominate and thence compute their on-shell action. For simplicity we will 
consider Einstein-Hilbert gravity here; generalizations to other classical gravitational theories follow along the lines of \cite{Dong:2013qoa,Camps:2013zua}.

To enforce the boundary conditions in the gravitational solution, we examine the metric close to $\fixM_q$. Consider a wave equation in the local coordinates of  \eqref{eq:lm1}. It is easy to see that it admits four local mode solutions, viz.,  $(r^q \, e^{i\,\tau})^{\pm \omega }$ and $(r^q \, e^{i\,\tau})^{\pm i\,\omega\, }$ in the vicinity of $r=0$.  To ascertain which of these is admissible and thus give explicit boundary conditions, we invoke two facts.
Firstly, the replica symmetry requires a $2\pi$ periodicity for fields as functions of $\tau$, restricting us to purely oscillatory functions and thereby fixing $\omega \in {\mathbb Z}$.  Secondly, regularity of the covering space implies that the fields have to admit an expansion in powers of $r^q\, e^{\pm i\,\tau}$, thus  preventing fields from diverging  at $r=0$ for integer $q$. Combining these facts we learn that $r^q e^{\pm i \tau}$ will be the generic behaviour of the metric near the origin.\footnote { The astute reader may worry that as a consequence we will have some components of the curvature being singular near $r=0$ for $q\notin {\mathbb Z}$. This, while true, turns out to be tamable -- the singularities will be integrable in a suitable sense, as we shall see.}

From the above discussion we then learn that the most general ansatz for the geometry near $\fixM$ compatible with our boundary conditions is:\footnote{ Notation: Greek (lowercase) indices refer to the full spacetime, mid-alphabet lowercase Latin indices $i,j,\cdots$ refer to the tangent space of the fixed point set $\fixM_q$, and early-alphabet lowercase Latin indices $a,b,\cdots$ refer to the normal bundle of $\fixM_q$.}
\begin{multline}
ds^2 =
         \left(q^2 dr ^2+r ^2 \, d\tau^2\right)+ 
        \left(\gamma_{i j}+2\, K_{ i j}^x\, r^q\cos \tau+2\, K_{ i j}^t\, r^q \sin \tau \right) dy^i \, dy^j  
         \\
        +\left[r^{f_q(q-1)}-1\right]
         \delta g_{\mu\nu}\, dx^\mu\,dx^\nu +\cdots   \,,
\label{eq:lm2}
\end{multline}
where $f_q$ is some analytic function of $q$  such that $f_q(q-1)$ takes nonnegative even integer values when $q$ is a positive integer. This metric is smooth and $\Zn$ symmetric for integer $q$.  Evaluating the Ricci tensor for the geometry \eqref{eq:lm2} near $q=1$, we find divergent contributions proportional to  $(q-1)\, \frac{K^a}{r}$ where $K^a \equiv K^a_{ij}\,\gamma^{ij}$ is the trace of the extrinsic curvature.  This divergent contribution cannot be compensated by modifying other components of the metric. We are thence led to conclude that the equations of motion give us a constraint on the allowed $\fixM_q$. The allowed codimension-2 surfaces are required to have vanishing trace of the extrinsic curvature in the normal directions. Since we have a $t \rightarrow -t$ symmetry, we have trivially $K^t=0$.
The constraint is equivalent to the minimal surface condition of \cite{Ryu:2006bv}: 
\begin{equation}
\lim_{q\to 1} \,\fixM_q  \to \extr \,, \qquad \extr \in \bulk \; \text{with}\;\;  t=0, \;K^x =0\,.
\label{eq:lmrt}
\end{equation}  

Let us record here for completeness that we have included in our metric ansatz \eqref{eq:lm2} potential contributions from first subleading orders (the $\delta g$ term).\footnote{ We should note that such terms are sometimes desired in order to cancel subleading divergences in higher derivatives theories \cite{Camps:2014voa}.} It may be verified that this term by itself cannot cancel the divergences arising from the $r^{q}$ terms in the metric.

At this point we are almost done: once we know that one should restrict to $\extr$ as in \eqref{eq:lmrt} all that is left is to compute the on-shell action of this geometry. It turns out to be convenient to compute not the action itself, but its $q$-derivative $\partial_q I[\hat{\bulk}_q]$ as explained around \eqref{eq:renyiqder}. This perspective was explained in \cite{Lewkowycz:2013nqa} and is based on the covariant phase space approach used in the black hole context \cite{Iyer:1994ys}. As shown in \cite{Dong:2016fnf}, there is in fact a simple geometric prescription for this object $\partial_q I[\hat{\bulk}_q]$ for any value of $q$, so we do not have to set $q=1$ in the following discussion. 

The key point is to view $\partial_q$ as a variation of the bulk solution (and its boundary conditions). Standard variational calculus says that any variation of the action can be written as a combination of the equations of motion and boundary terms (using integration by parts where necessary):
 \begin{equation}
 \delta I[\hat{\bulk}_q]=\int_{\bulk_q} \left[\text{EOM}\cdot \delta g_q +d \Theta(g_q,\delta g_q) \right] \,.
 \label{eq:var0}
 \end{equation}
For a typical variation that appears in a standard AdS/CFT calculation, this would evaluate to a term at the asymptotic boundary $\partial \bulk_q =\bdy_q$. However,  we wish to consider the variation of $q$, which instead changes the boundary condition near the fixed point set $\fixM_q$. For the choice $\delta g=\partial_q g$ the variation satisfies $\partial_q g_q \big|_{\bdy_q}=0,\partial_q g_q\big|_{\fixM_q} \not = 0$. So we see that the change of the action engendered by the replica index variation is localized at the fixed point locus and has no contribution from the asymptotic boundary of the spacetime. One may therefore write
 \begin{equation}
 \partial_q I[\hat{\bulk}_q]=\int_{\fixM_q(\epsilon)} \Theta(g_q,\partial_q g_q)
 \label{eq:var1}
\end{equation} 
 where we have chosen to regulate the result by blowing up the singular locus to a tubular neighbourhood. In other words the fix point set $\fixM_q$ which was at $r=0$ is now being regulated by a codimension-1 surface 
 $\fixM_q(\epsilon)$ at $r=\epsilon$. We will obtain the correct answer when   $\epsilon\to 0$.

In the present case we will not actually evaluate this integral (which can be done given the symmetries),  but will follow an equivalent route. In the presence of a boundary for the variational calculus to be well-defined and give the correct equations of motion, we would need to supply the correct boundary terms. While in our case  the surface $\fixM_q(\epsilon)$ is not really a physical boundary, one may for purposes of evaluation imagine that it is and ascertain the corresponding boundary terms. The advantage of this trick is that the on-shell action will be given simply by evaluating these contributions. For Einstein-Hilbert  gravitational dynamics we evaluate the Gibbons-Hawking contribution from $\fixM_q(\epsilon)$\footnote{ To prevent notational clutter, we drop the integration measure for simplicity in the formulae henceforth. The measure should hopefully be clear from the context.}
\begin{equation}
\partial_q I[\hat{\bulk}_q]=-\partial_q I_{bdy}[\hat{\bulk}_q]
\,, \qquad 
I_{bdy}[\hat{\bulk}_q]=\frac{1}{8 \pi G_N}\int_{\fixM_q(\epsilon)} \, {\cal K}_\epsilon
\label{eq:var2}
\end{equation}
where ${\cal K}_\epsilon$ is the trace of the extrinsic curvature of the codimension-1 surface $\fixM_q(\epsilon)$, evaluated with the outward pointing normal vector. Again, this holds in the $\epsilon\to0$ limit.

Working in the local coordinates \eqref{eq:lm2} in an open neighbourhood of $\fixM_q$, one finds  ${\cal K}_\epsilon=\frac{1}{q \,\epsilon}$, and thus we get\footnote{ The variation of the metric \eqref{eq:lm2} at 
$\fixM_q$ is $g^{r r} \partial_q g_{r r} \big|_{\fixM_q}=\frac{2}{q}$ and vanishes for the other components.} for the modular entropy:
\begin{equation}
\partial_q I[\hat{\bulk}_q]= \frac{\text{Area}(\fixM_q)}{4\, q^2 G_N}
\label{eq:renderq}
\end{equation}
which as $q \rightarrow 1$ gives us the RT formula.

Before moving to the Lorentzian case, some words of caution are in order. The orbifold picture allows us to analytically continue the on-shell action $I[\bulk_q]$ to non-integer $q$. The physical interpretation of the (parent space) solution for non-integer $q$ is unclear, but these geometries are just an intermediate step to compute the action.  In this way, even if for integer $q$ these geometries do not have singularities, some components of the Riemann tensor will go like $\sim r^{q-2}$ for $q \notin {\mathbb Z}$. However, these components neither appear in the equations of motion, nor the evaluation of the action, and are thus mostly harmless.

%~~~~~~~~~~~~~~~~~~~~~~~~~~~~~~~~~~~~~~~~~~~~~~~
\subsection{Covariant generalization to time-dependent situations}
\label{sec:covgen}
%~~~~~~~~~~~~~~~~~~~~~~~~~~~~~~~~~~~~~~~~~~~~~~

As we have extensively presaged in the earlier sections, in genuine time-dependent circumstances, we cannot invoke the trick of passing to a path integral over a Euclidean manifold.\footnote{ In the absence of time-reflection symmetry, the analytic continuation of $t \to i\,\tE$ will lead to a complex manifold. Furthermore, analytic continuation to Euclidean signature would not work in generic cases where the time-dependent sources are non-analytic.} 
Rather, we have a Schwinger-Keldysh or time-folded path integral which has to be dealt with in Lorentzian spacetime. Indeed the field theory construction of the density matrix explained in \S\ref{sec:qft} requires evolution from the initial state up until the moment of interest, say $t$, and then retracing one's footsteps back to the far past. This forward-backward evolution induces a kink at the Cauchy slice $\CSA\equiv \regA \cup \regAc$  on the boundary $\bdy$, as we only retain the part of the geometry to its past, i.e.\ $J^-[\CSA]$.

\begin{figure}[htbp]
\begin{center}
\includegraphics[width=5in]{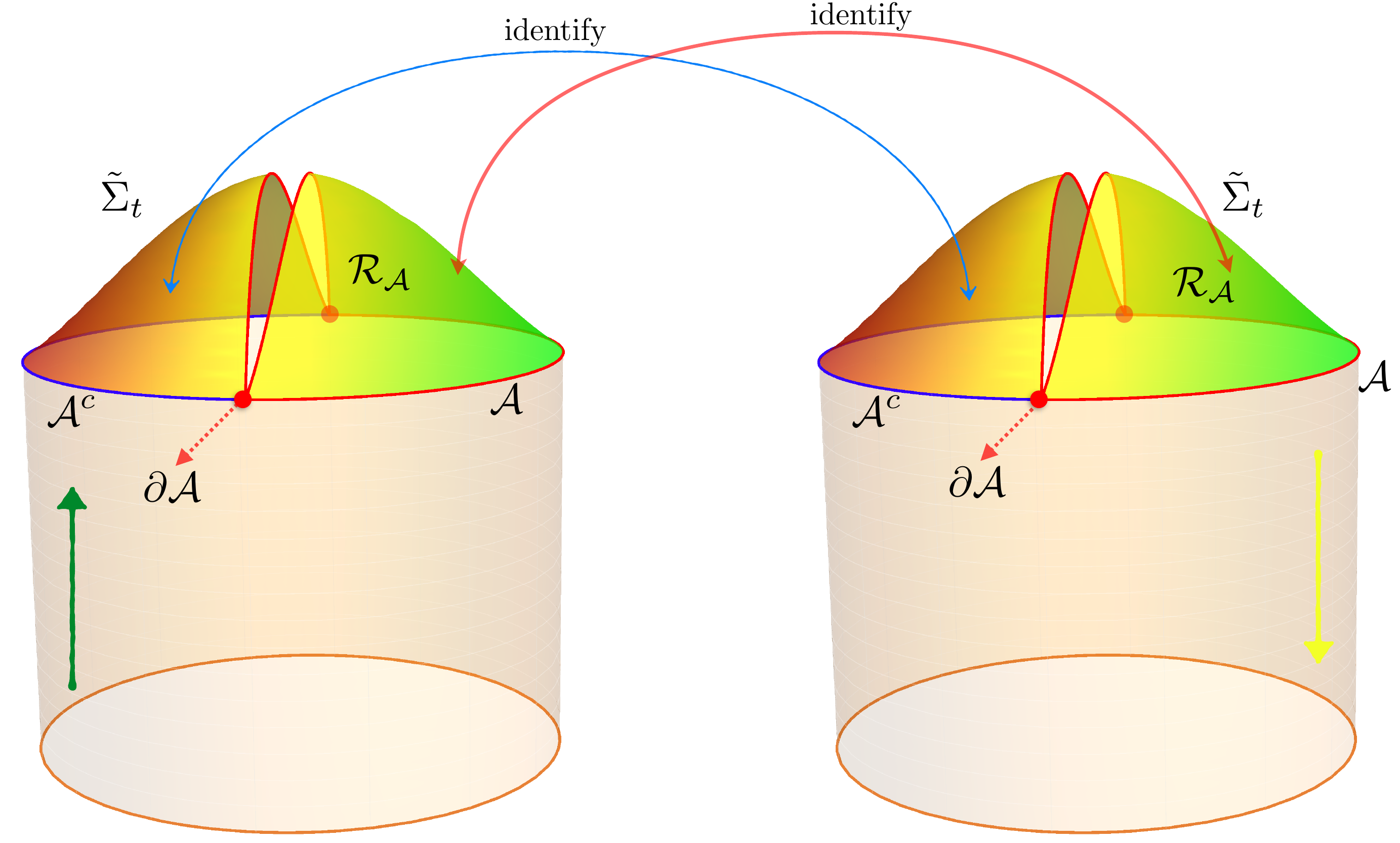}
\caption{The bulk construction of the trace of the reduced density matrix involves two copies of the spacetime in question, which are glued across the part $\homsurfAc$ of the bulk Cauchy slice associated with $\regAc$. Taking the trace corresponds to gluing across the part $\homsurfA$ of the bulk Cauchy slice associated with $\regA$. }
\label{fig:bulksk}
\end{center}
\end{figure}

In the bulk there will be an analogous fold along some Cauchy slice $\bulkCS$, with the proviso that the bulk evolution will proceed only in the part of the spacetime to the past of $\bulkCS$, i.e., in $\bulkJ^-[\bulkCS]$.\footnote{ We will use a tilde to distinguish bulk Cauchy surfaces and causal sets  from analogous quantities on the boundary.} In other words the initial conditions are evolved forward from $t=-\infty$ up to $\bulkCS$ and then we evolve back to construct the bulk Schwinger-Keldysh contour.\footnote{ Similar statements apply to states prepared at $t=t_0 <0$, for instance by a Euclidean path integral, and then evolved up to $t=0$ perhaps with sources, etc.} This forward-backward evolution 
through $\bulkCS$, across which two copies of the bulk manifold are glued together, is illustrated in Fig.~\ref{fig:bulksk}. On the Cauchy slice as we   reverse the evolution,  we have to provide appropriate matching (boundary) conditions\footnote{ These boundary conditions guarantee that we have a well defined variational principle and that the variation of the action doesn't contain boundary terms at $\bulkCS$.} .

% Figure 

 As described in \S\ref{sec:qft}, we may move $\CSA$ itself as long as $J^-[\entsurf]$ is unmodified. Let us say,  for  definiteness, we make a particular choice  and stick with it w.l.o.g. This however does not single out a bulk Cauchy slice,  since the boundary time coordinate does not extend uniquely into the bulk. As remarked in \S\ref{sec:intro}, $\bulkCS$ can lie anywhere in the Wheeler-DeWitt patch (part of the bulk spacelike to $\CSA$) provided $\partial \bulkCS = \CSA$.  For the moment we will take $\bulkCS$ to be some representative in this Wheeler-DeWitt patch (as shown for instance in Fig.~\ref{fig:bulkDoms}); we will learn in the course of our analysis of potential restrictions on  bulk Cauchy slices. 

Basically, we are extending the construction of the field theory into the bulk in an intuitive manner.  Each piece of the bulk glued across $\bulkCS$ corresponds to the state $\ket{\psi}$ (forwards) or its conjugate $\bra{\psi}$ (backwards) in the gravitational description. This extension of the field theory Schwinger-Keldysh contour allows for computation of holographic real time (in-in) correlation functions \cite{Skenderis:2008dg}.

Having understood how to set up the Lorentzian problem in gravity, we now have to face up to the harder question:  ``What is the dual of tracing out degrees of freedom in $\regAc$?'' We will first analyze the kinematic part of the construction and pick out the boundary conditions involved in computing  $\tr (\rhoA)^q$. We will then argue that the trace over $\regAc$ can only be done if $\bulkCS$ satisfies some specific properties.\footnote{ It is worth contrasting this with the computation of correlation functions for which any $\bulkCS$ in the Wheeler-DeWitt patch of $\CSA$ is acceptable \cite{Skenderis:2008dg}.}

It bears emphasizing that our construction has two distinct components: 
\begin{itemize}
\item[(i).] constructing the dual of  field theory density matrix $\rho(t)$.
\item[(ii).] understanding the process of tracing out $\regAc$.
\end{itemize}
The first involves the Schwinger-Keldysh framework, while the second involves identifying the part of the bulk spacetime we trace over. The latter issue is already present in the time-reflection symmetric case, should we  view the discussion of \S\ref{sec:lmreview} in Lorentzian terms.

%~~~~~~~~~~~~~~~~~~~~~~~~~~~~~~~~~~~~~~~~~~~~~~~
\subsubsection{Kinematics: setting up the bulk construction}
\label{sec:kinL}
%~~~~~~~~~~~~~~~~~~~~~~~~~~~~~~~~~~~~~~~~~~~~~~

Let us first try to set up the gravitational problem and come up with an ansatz which we can use to explore the bulk solution of relevance. From an operational point of view, we need to understand how to translate the complicated path integral construction of the boundary field theory in terms of the bulk variables. The upshot will be the following: we will first ascertain what it takes to compute the R\'enyi entropies at integer $q>1$ in the bulk. Subsequently, we will argue that the computation can equivalently be done upon taking a ${\mathbb Z}_q$ quotient in a `single fundamental domain', which  has the same asymptotics as the  $q=1$ geometry but differing boundary conditions. It is the latter that implies the existence of a fixed point locus $\fixM_q$ on $\bulkCS$.

The R\'enyi entropy, $\tr_\regA (\rhoA)^q$, is constructed in the boundary from $q$ copies of the density matrix itself on $\CSA$. This has  $q$ forward time-contours (or ket-folds) and $q$ backward time-contours (or bra-folds) glued together appropriately. The gluing is essentially dictated by the split  
$\CSA = \regA \cup \regAc$ which defines   $\tr_\regA (\rhoA)^q$.   Concretely, the part corresponding to $\regA$ is glued across from one density matrix to its immediate neighbour, while that corresponding to $\regAc$ is glued back onto itself.

\paragraph{1. Local Rindler coordinates in the bulk, for $q=1$:} Consider the $q=1$ geometry  used to construct the bulk analogue of $\tr_\regA (\rhoA)$ (equivalently $\tr( \rho(t))$). This geometry corresponds to two $\bulkJ^-[\bulkCS]$ segments glued across at $\bulkCS$. A priori, it is not clear how the different regions $\regA,\regAc$ are encoded in the bulk. We will take a cue from the boundary and assume that there is a simple extension of the split of the boundary Cauchy slice into the bulk; even if this seems ad-hoc, the existence a codimension-2 surface implementing this bipartitioning will be a consequence of the bulk replica symmetry, as we will discuss later. We start with an ansatz that we can divide the bulk Cauchy slice into two regions $\bulkCS=\homsurfA \cup \homsurfAc$, which intersect along a bulk codimension-2 surface $\fixM$. This will be the bulk extension  of  $\entsurf$; see Figs.~\ref{fig:bulkDoms} and \ref{fig:bulksk}. While we do not know (as yet) how to determine $\fixM$, we can ascertain how the boundary statements are reflected in the bulk given such a demarcation. We will later see that $\fixM$ is fixed by analytic continuation of the $\Zn$ symmetric fixed points for the replicated geometries.

\begin{figure}[h!]
\begin{center}
\vspace{5mm}
\includegraphics[scale=0.5]{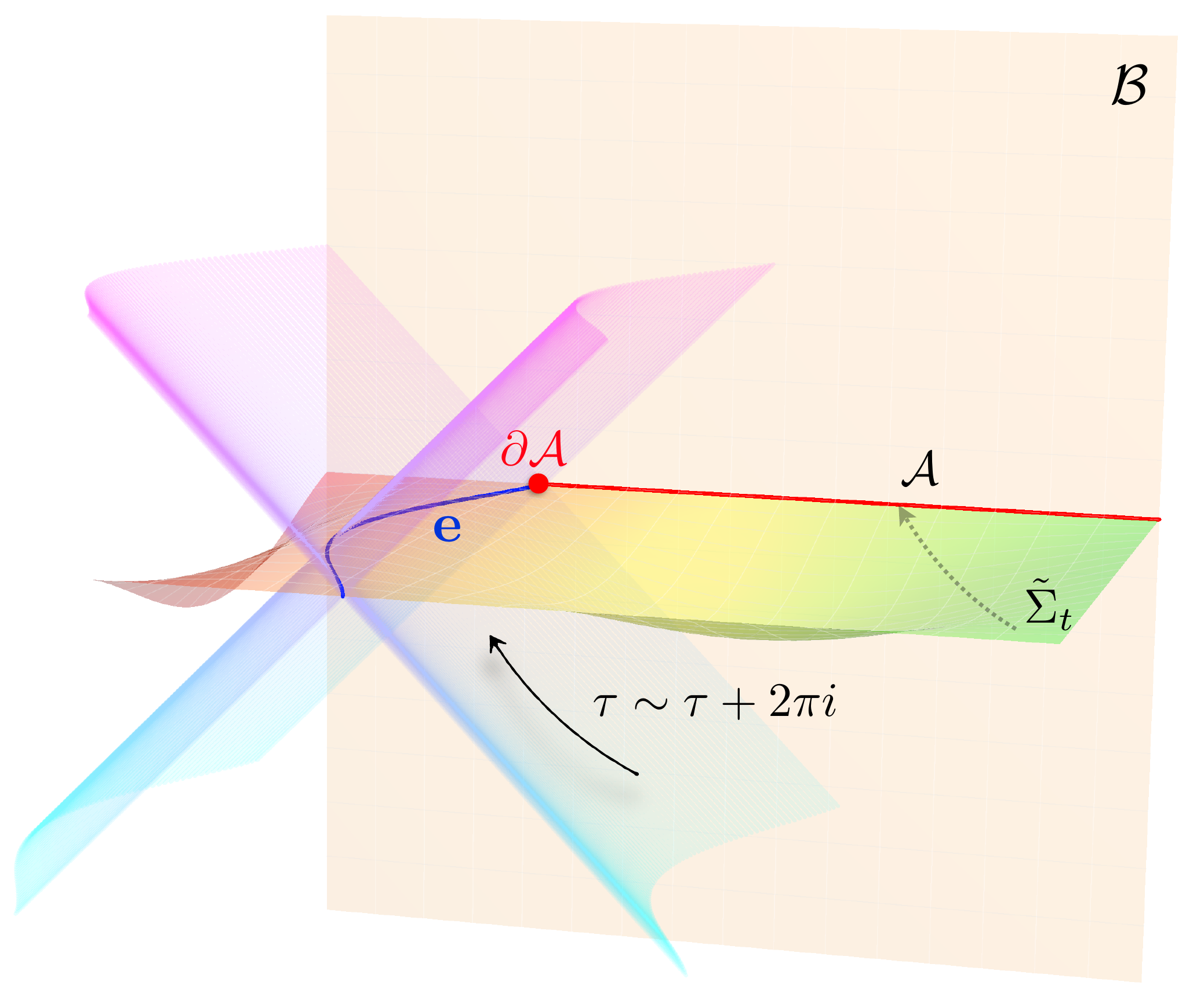}
\vspace{5mm}
\caption{A sketch of the local neighbourhood of the singular locus $\fixM$ ending on the boundary at $\entsurf$. We show the local horizons of this surface and the boundary conditions imposed as we cross the Cauchy slice $\tilde{\Sigma}_t$. The meaning of the identification $\tau \sim \tau + 2\pi i$ is explained in the main text.}
\label{fig:bulkbcs}
\end{center}
\end{figure}

 Of particular utility will be to understand how to extend the boundary coordinates into the bulk. In particular, we should find an analogue of the  $\tau$ coordinate which helps identify the boundary conditions. To get a sense of how to proceed, we can foliate the causal development of $\homsurfA$ in Rindler-like coordinates. We will use the fact that one can naturally write the metric close to any codimension-2 surface (such as $\fixM$) as:
\begin{equation}
ds^2 = dr^2 - r^2 \, d\tau^2 + \left(\gamma_{mn} + r\, e^{\pm \tau} \,
K^\pm_{mn} + \cdots \right) dy^m\, dy^n\,.
\label{}
\end{equation}  
This metric is completely regular since we are just rewriting the original metric in adapted coordinates. 

However, it would be useful to view this slightly differently. For the Rindler-like observer one encounters four horizons emanating from the codimension-2 surface; see Fig.~\ref{fig:bulkbcs}. As explained in \S\ref{sec:qft}, in these coordinates a horizon crossing can be understood as $\tau \rightarrow \tau + i
\frac{\pi}{2}$ (along with $r \to i^{-1} r$). The complex shift is a useful mnemonic to remember the boundary conditions; the passage to complex values of $\tau$ should not be viewed as a fundamental necessity.\footnote{ To reiterate, this can be clearly be seen in the Cartesian $\{x,t\}$ coordinates, for there it is a simple swap $x\leftrightarrow t$.}  It captures the local geometry in a neighbourhood of $\fixM$ efficiently. Regularity 
demands that one should go back to the starting Rindler wedge (say the domain of development of $\homsurfA$) after crossing four horizons. This is what we propose to encode as  $\tau \sim \tau+2\pi i$. This is the Lorentzian analogue of the fact that we have a Euclidean time circle with an appropriate size.

A more physical way to state the boundary conditions is the following: in $\regA$ and thence in $\homsurfA$, fields that approach $\entsurf$ and $\fixM$, from within $\regA$ and  $\homsurfA$ respectively, should behave like local Rindler modes with an effective temperature $\frac{1}{2\pi}$. This  local monodromy condition calls for the choice $\tau \sim \tau + 2\pi i$ above; it is only in this restricted sense we talk about complex shifts of the temporal coordinate. 
 
If we have $k=1,2,\cdots , q$ domains in the bulk with coordinates $\{ r_k ,\tau_k, y^m_k \}$, then we would have a similar story  in each replica copy. The equivalent boundary conditions would be to essentially demand that every time four horizons are crossed one goes to the next replica: $\tau_k \sim \tau_{k-1}+2\pi i \,, \;
 r_k \sim r_{k-1}\,, \; y^m_k \sim y^m_{k-1}$.  Given the assumption of bulk replica symmetry  and that the boundary fixed point $\entsurf$ extends naturally into a bulk fixed point  $\fixM$, it seems natural to think of the bulk as having this branched cover structure.  We should also understand under what conditions the metric is a smooth solution to the equations of motion. 

\paragraph{2. Construction of the replicated geometries:} 

With this in mind, let us move on to the construction of the bulk geometry $\bulk_q$ dual to $\tr (\rhoA)^q$. In the boundary, we have $q$ copies of a folded geometry glued cyclically along part of $\CSA$. We will now assume that the partition function on this geometry $\bdy_q$  can be computed in the saddle point approximation by a bulk geometry similar to those of  \cite{Skenderis:2008dg}.  As in the Euclidean case, we are going to assume that the boundary replica ${\mathbb Z}_q$ symmetry extends into the bulk. In this process, the boundary fixed point set $\entsurf$ also gets continued along some bulk codimension-2 locus of fixed points, which we will denote as $\fixM_q$.\footnote{ If there is no bulk fixed point, we expect that the entropy is zero, as discussed in \cite{Lewkowycz:2013nqa}} In each replica, this fixed point locus by construction lies on $\bulkCS$, a Cauchy slice whose boundary is $\CSA$, and serves to demarcate the surface into two parts. Given these, it seems natural to expect that this requires the boundary branched cover structure inherent in the replica construction to be inherited by the holographic map in the bulk. In this way, we have a branched geometry, whose action should correspond to $\tr (\rhoA)^q$. We also have a well-defined codimension-2 surface which extends $\partial A$ to the bulk.  In a suitable limiting sense, as described below, one should think of $\fixM$ above as  $\fixM_{q \rightarrow 1}$.

\begin{figure}[htbp]
\begin{center}
\includegraphics[width=5.5in]{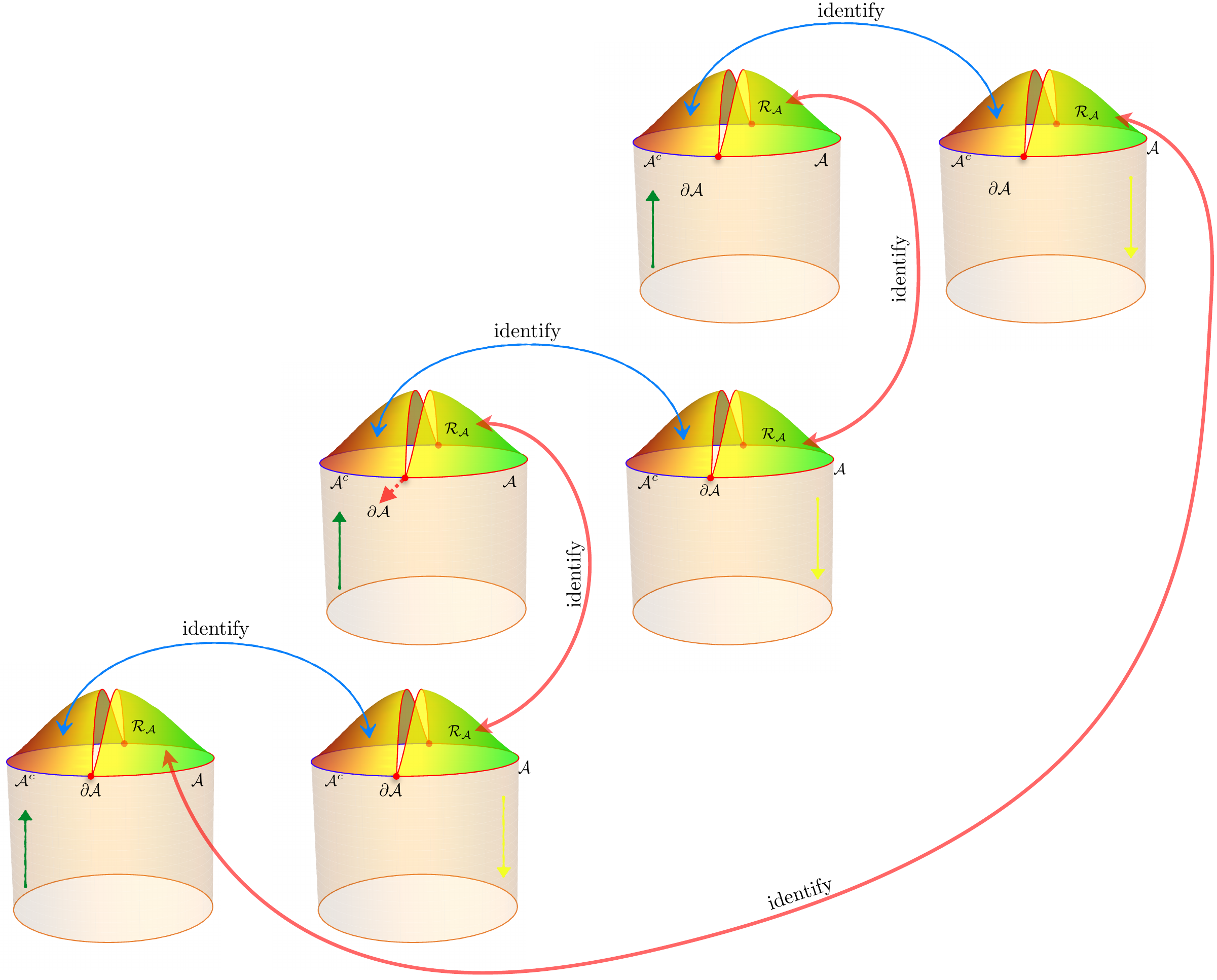}
\caption{The construction of the third R\'enyi entropy using the holographic map. We need six copies of the spacetime to construct the replicated Schwinger-Keldysh path integral. The identifications of $\homsurfAc$ are as described before in Fig.~\ref{fig:bulksk}, while the identifications of $\homsurfA$ are across the replica copies cyclically.}
\label{fig:bulkrhoA3}
\end{center}
\end{figure}

  The previous discussion can be heuristically viewed as follows: we divide $\bulkCS$ into two pieces across $\fixM_q$ as indicated and glue the components differently across the multiple copies of the density matrix viewed as geometries. That is, we now require that fields which approach $\bulkCS$ inside $\homsurfA$ on the $k^{\rm th}$ copy, pass onto the $(k+1)^{\rm st}$ copy as depicted for example in Fig.~\ref{fig:bulkrhoA3}. This cyclic gluing condition can equivalently be phrased as saying that the fields which approach $\fixM_q$ through $\homsurfA$ feel a local temperature $\frac{q}{2\pi}$; this constrains the mode functions for the fields.\footnote{  In addition it determines the initial conditions for evolving fields from one side to the other side of the Rindler horizons (of each copy) for it gives appropriate matching conditions for the modes across the horizons.} These should be thought of as prescribing the relevant boundary conditions for our problem in gravity, thus naturally extending the field theory discussion of \S\ref{sec:qft}; cf., Fig.~\ref{fig:bulkbcs}. Of course, given these boundary conditions, the dual geometry will be rather complicated. For one, it will be very different from the off-shell picture of \cite{Fursaev:2006ih}  and for another, for integer $q>1$, we will not be able to say much more beyond the simplest cases.  

We have argued that $\fixM_q \subset \bulkCS$, but since $\bulkCS$ was a priori any bulk Cauchy slice inside the Wheeler-DeWitt patch, we seem to have a large amount of freedom in the location of the fixed point set. One can fix some of this ambiguity by invoking the causality properties satisfied by entanglement entropy to argue that the fixed point set $\fixM_q$ must lie in the causal shadow of $\domdA$ \cite{Headrick:2014cta}.
This does not pin down the surface in any way, for the causal shadow of a boundary domain of dependence generically is a bulk codimension-0 volume.

We have basically distilled the construction of $\bulk_q$ (similar to the Euclidean case) to focus our attention on only one copy $\hat{\bulk}_q \equiv \bulk_q/\Zn$ of the replica geometry. This geometry has a ${\mathbb Z}_q$ fixed point set which, being spacelike, could be called an ``instanton conical singularity''.  More precisely it should be considered to be a codimension-2 S-brane in spacetime \cite{Gutperle:2002ai} with tension set by $q$. Our task is now to understand what the constraints are on $\fixM_q$ (and also $\bulkCS$) that respect the boundary conditions of our problem. Thus having used up the kinematic data to motivate an ansatz for the gravitational construction, we now appeal to the dynamics of the bulk gravitational theory to provide constraints on $\fixM_q$, using the fact that the covering space needs to satisfy the correct boundary conditions and equations of motion near the fixed points.

We have generalized the discussion of \cite{Skenderis:2008dg} beyond boundary contours where the gluing doesn't have  spatial dependence since the gluing conditions for the R\'enyi entropies are different on the two sides of the entangling surface. We have done this by extending the boundary contour into the bulk in the most straightforward manner  that is compatible with the replica symmetry: gluing purely Lorentzian segments and imposing the proper boundary conditions. This is a natural extension of \cite{Skenderis:2008dg} which we assume without further justification in what follows.

One of the features of the Schwinger-Keldysh construction is a  redundancy built into the construction. This can be understood from the ability to implement field redefinitions in the doubled theory, cf., \cite{Haehl:2015foa}. This allows certain deformations of the contour which nevertheless end up giving the same physical answers for observables (including the on-shell action). Readers may be familiar with a related statement in thermal field theory, where there is a one-parameter family of  Schwinger-Keldysh contours, characterized by the two Lorentzian contours separated by an arbitrary Euclidean distance, with the proviso that the total contour be  periodic in imaginary time with period $\beta$. Though this argument typically relies on the analyticity of thermal correlators, we cannot rule out in general a deformed contour in the bulk which computes the R\'enyi entropies of interest.\footnote{ The future gluing condition in the Schwinger-Keldysh contour is effectively a projection of the final density matrix against the maximally entangled state in the doubled system. The latter is obtained as the $T\to \infty$ limit of the thermal density matrix. We thus can imagine a deformation wherein we glue a copy of the Euclidean instanton corresponding to this limiting solution.}  These may perhaps have additional Euclidean segments, but the general expectation is that they will also have the same on-shell action as the configuration that we favour with minimal Euclidean excursions (just those necessary for a correct $i\varepsilon$ prescription). It would be interesting to examine this issue further.

%~~~~~~~~~~~~~~~~~~~~~~~~~~~~~~~~~~~~~~~~~~~~~~~
\subsubsection{Dynamics: equations of motion and extremal surfaces}
\label{sec:dynL}
%~~~~~~~~~~~~~~~~~~~~~~~~~~~~~~~~~~~~~~~~~~~~~~

In \S\ref{sec:kinL}, we have used the kinematic data at hand to set up the problem. When all the dust has settled, we have essentially reduced our attention to a fundamental domain $\hat{\bulk}_q$ of the bulk under the replica $\Zn$ symmetry, namely a Schwinger-Keldysh double geometry constructing the dual of the trace of the total density matrix $\tr \rho(t)$ with a $\Zn$ symmetric fixed point set, $\fixM_q$, localized on the Cauchy surface $\bulkCS$. The remaining task at hand is to employ the bulk equations of motion, see what they imply for $\fixM_q$, and compute the on-shell action thereafter.

\paragraph{1. The extremality condition:}
We have described the boundary conditions that we need to satisfy in \S\ref{sec:kinL}. As in \S\ref{sec:lmreview} it is  useful to switch to Rindler-like coordinates $\{r,\tau\}$ for the normal bundle of $\fixM_q$ in the bulk.  In the following discussion, we will focus on the forward segment of the Schwinger-Keldysh contour ($\tau<0$).\footnote{ Since the analysis is local below $\bulkCS$, we do not need to worry about the kink.} Analogous to \eqref{eq:lm2} in \S\ref{sec:lmreview}, the metric is constrained by the $\Zn$ symmetry, boundary conditions, and regularity for integer $q$ to have the following expansion in the vicinity of $\fixM_q$:
\begin{multline}
ds^2 = \left(q^2 dr ^2-r ^2 \, d\tau^2\right)+ 
        \left(\gamma_{i j}+2\, K_{ i j}^x \,r^q 
        \cosh \tau+2\, K_{ i j}^t \,r^q 
        \sinh \tau\right) dy^i \, dy^j \\ +\left[r^{f_q\,(q-1)}-1\right]
         \delta g_{\mu\nu} \, dx^\mu\, dx^\nu +\cdots  
\label{eq:lmL}
\end{multline}
where we denote the coefficients of the $r^q$ terms as $K^a_{i j}$ because in the $q \rightarrow 1$ limit they give the extrinsic curvature.

With this ansatz for the geometry, we can now analyze the consequences of the equations of motion. This is in fact quite easy, since the local geometry resembles the Euclidean discussion. We have a deviation away from flat space (in Rindler coordinates) owing to an instantonic brane source with tension set by $q$. The gravitational equations of motion away from $\fixM_q$, however, do not care about this.

Indeed, evaluating the terms in the equations of motion for Einstein-Hilbert dynamics in the bulk, we find potentially divergent terms proportional to %
\begin{align}
\text{EOM}^a \propto \frac{q-1}{r} \, K^a  + \text{regular}^a
\label{eq:eomL}
\end{align}
for small $q-1$.  Basically, the presence of the extrinsic curvature terms in \eqref{eq:lmL} leads to potentially singular behaviour of the Ricci tensor in the neighbourhood of $r =0$. These cannot be compensated for by any correction to the metric that respects the $\Zn$ symmetry and boundary conditions. 
 
One then learns that the trace of the extrinsic curvature in each of the normal directions must vanish, i.e., $K^t = K^x =0$. While this statement refers to the trace in the timelike ($K^t$) and spacelike ($K^x$) directions respectively, we can by taking suitable linear combinations express this in terms of the null expansions which are more natural for codimension-2 spacelike surfaces in  Lorentzian manifolds. Defining $x^\pm = \frac{1}{\sqrt{2}}\, \left(x^0 \pm x^1\right)$ we thus have the extremal surface condition postulated in \cite{Hubeny:2007xt}, viz.,\footnote{ Note here that $K^0_{ij}$ is the component of the extrinsic curvature in the timelike normal direction to a codimension-2 surface (likewise $K^1_{ij}$ is the corresponding spacelike component) and should not be confused with the extrinsic curvature for $\bulkCS$ (which has a timelike normal), for which we use the symbol ${\cal K}$ when necessary.}
\begin{equation}
\begin{split}
K^a =0 & \;\; \Longrightarrow \;\; \theta^\pm = \frac{1}{\sqrt{2}} \left(K^0 \pm K^1\right) = 0 \,, \\
& \;\; \Longrightarrow \;\;  \lim_{q\to 1} \fixM_q = \extr \,, \qquad \extr \in \bulk \;\text{is extremal}.
\end{split}
\label{eq:extremalEom}
\end{equation}  
Having ascertained the dynamical constraint on $\fixM_q$ in the limit $q\to1$, let us return to our earlier discussion. We originally argued in \S\ref{sec:kinL} that $\fixM_q$ should, by virtue of the replica symmetry assumption, lie on the Cauchy surface $\bulkCS$ which we pick to construct the density matrix $\rho(t)$ for the entire system. As indicated in that context, the choice of $\bulkCS$ is restricted by the fact that it be spacelike to $\CSA$ and $\partial \bulkCS = \CSA$, but is otherwise unconstrained.  However, the dynamics indicates that not all such $\bulkCS$ would be acceptable in semiclassical saddle point solutions to the gravitational path integral. While an arbitrary $\bulkCS$ in the Wheeler-DeWitt patch of the boundary Cauchy surface may be used a priori to construct  $\tr \rho(t)$, the semiclassical saddle point of the Lorentzian path integral for $\tr(\rhoA^q)$ (near $q \sim 1$) only chooses those that pass through the extremal surface, see Fig.~\ref{fig:bulkDoms}. More pertinently, we conclude that $\tr(\rhoA^q) $ can be constructed by the Lorentzian prescription provided $\extr \subset \bulkCS$. This restriction does not originate from the general Schwinger-Keldysh construction, but rather is specific to the process of tracing  out the degrees of freedom in $\regAc$. More explicitly, it originates from the fact that we are effectively introducing a singularity along $\fixM_q$.

\paragraph{2. The on-shell action:} The computation of the on-shell action, once we realize that the fixed point locus of replica symmetry becomes the extremal surface in the $q\to1$ limit, proceeds in a similar manner as before, modulo a few small subtleties.  The main difference is the fact that we have to work directly in Lorentzian signature, which means that the regulated codimension-1 surface $\fixM_q(\epsilon)$ would be more complicated. We will additionally have to deal with the presence of light-cone singularities in \eqref{eq:lmL}.  For instance when $q \notin {\mathbb Z}_+$, some curvature components behave as $r^{q-2}$. The functional form is similar to the Euclidean case, but now the origin of the normal plane to $\fixM_q$ is blown up in Lorentzian signature to a codimension-1 null surface, which is the lightcone emanating from the origin. Fortunately, these turn out to be mild singularities which do not contribute to the evaluation of the action.

The non-trivial computation here is that of the R\'enyi entropies, which are technically more challenging than in the Euclidean case. We have found it useful to compute the quantity $\tilde{S}^{(q)}_\regA$ introduced in \eqref{eq:renyiqder} directly, but even this requires careful handling of an $i\varepsilon$ prescription. We demonstrate in Appendix~\ref{sec:action} that this can in principle be done and provide a few simple examples there. Presently we will give a sketch of how such a computation proceeds.

Assuming that the extremal surface arises as a consequence of a well-defined variational principle as in \eqref{eq:var1}, all that remains is to compute the boundary term. As before the computation requires us to evaluate the Gibbons-Hawking term for Einstein-Hilbert gravitational dynamics, cf., \eqref{eq:var2}
\begin{equation}
\partial_q I[\hat{\bulk}_q]= -\frac{1}{8 \pi G_N} \; \partial_q \; \int_{\fixM_q(\epsilon)} \; {\cal K}_\epsilon \,.
\end{equation}
We can proceed thus far without worrying about the change in the signature of the metric. Now we have however to face up to the fact that the codimension-1 regulator surface $\fixM_q(\epsilon)$ defined as the hypersurface $r=\epsilon$ comprises four distinct segments (two spacelike and two timelike). The computation has to be done from scratch, because even under analytic continuation this surface does not give us the $r=\epsilon$ locus for the Euclidean problem in \cite{Lewkowycz:2013nqa}. Note that the boundary terms at the Cauchy surface $\tilde{\Sigma}_t$ cancel out due the boundary conditions inherent in the prescription of \cite{Skenderis:2008dg}.

Despite these subtleties the evaluation of the boundary term works out to give the expected result for the covariant modular entropy:
\begin{equation}
\partial_q I[\hat{\bulk}_q]= i \frac{\text{Area}(\fixM_q)}{4\, q^2 G_N} \,.
\end{equation}
Thus we indeed obtain the area of the extremal surface as in \cite{Hubeny:2007xt} when we take the $q \rightarrow 1$ limit. Alternatively, the same result can be obtained by regularizing the singularity.

%~~~~~~~~~~~~~~~~~~~~~~~~~~~~~~~~~~~~~~~~~~~~~~~
\section{Discussion}
\label{sec:discussion}
%~~~~~~~~~~~~~~~~~~~~~~~~~~~~~~~~~~~~~~~~~~~~~~

We have now a derivation of the extremal surface prescription of \cite{Hubeny:2007xt} for computing holographic entanglement entropy in time-dependent states. We take the opportunity to comment on several consequences of this construction.

\paragraph{The homology constraint:}

 As explained elsewhere \cite{Haehl:2014zoa} the RT and HRT proposals for holographic entanglement entropy should respect the homology constraint. This requires that there be a spacelike codimension-1 interpolating homology surface, whose only boundaries are $\extr$ and $\regA$. The homology constraint is naturally incorporated in our construction.

 The boundary conditions relevant for computing the $q^{\rm th}$ R\'enyi entropy involves a cutting and gluing  in the bulk path integral. We have hitherto explained that our construction naturally restricts the HRT surface to lie on a bulk Cauchy slice $\bulkCS$ with $\bulkCS\big|_{_\bdy} \,= \CSA$. This in particular implies that $\extr$ splits $\bulkCS$ into two parts $\homsurfA$ and $\homsurfAc$ respectively with $\partial \homsurfA = \extr \cup \regA$ as required by the homology constraint. We need this to happen, since the powers of the reduced density matrix elements are obtained by cutting open the path integral along 
 $\homsurfA$ and sewing them cyclically. Specifically, we need to identify  $\homsurfA^{{\scriptsize (I)-}}$ with $\homsurfA^{{\scriptsize (I+1)+}}$ to respect the ordering of the matrix elements.\footnote{  One can also state the prescription more completely by requiring that we cut open along $\bulkCS$ and glue $(\homsurfAc)^{{\scriptsize (I)-}}$ is glued on back to $(\homsurfAc)^{\scriptsize{(I)+}}$.} In effect the basic construction singles out a bulk codimension-1 region $\homsurfA$ that serves to define how we carry out the trace over the degrees of freedom in the region $\regAc$. This picture should persist not just for $q\approx 1$ but even for finite $q$, in spite of the fact that the  corresponding geometry will be deformed significantly.

\paragraph{Entanglement wedge:} In our analysis we have started by fixing a boundary Cauchy slice $\CSA$ and picked a definite region on it. However, the computation of trace class observables is insensitive to the particularities of the slice; we are free to deform this as long as $J^-[\entsurf]$ remains unmodified \cite{Headrick:2014cta}. Picking various deformations of $\regA$ within its boundary domain of dependence $\domdA$, leaving the entangling surface untouched will in particular satisfy this requirement. 

The analogue of this freedom in the bulk corresponds to the choice of bulk Cauchy slices $\bulkCS$, which lie pinned at $\extr$. If we view $\homsurfA$, the piece of one such representative Cauchy slice as the bulk analogue of $\regA$ then as argued in \cite{Headrick:2014cta} the corresponding bulk domain is the {\em entanglement wedge} $\EWA = {\tilde D}[\homsurfA]$. 

In making these statements we are allowing ourselves the freedom to move $\regA$ into the future domain of dependence $D^+[\regA]$. Strictly speaking, in our analysis we have always chosen to fix $\CSA$ and $\regA$ {\em ab initio}; this would allow access only to the past domain of dependence on the boundary and correspondingly only the past half of the entanglement wedge in the bulk.

\paragraph{The ``dual" of tracing out:} We have traced out the degrees of freedom in the boundary to implement the replica trick. Attempting to do it similarly in the bulk, we have seen that one cannot do the replica trick on all Cauchy slices in the Wheeler-DeWitt patch. This suggests that a dual picture for tracing out boundary degrees of freedom exists only if the bulk Cauchy slice contains the extremal surface. 

This observation is important in the context of the subregion-subregion duality in holography. It has been argued by several authors \cite{Czech:2012bh,Wall:2012uf,Headrick:2014cta,Almheiri:2014lwa,Jafferis:2015del,Dong:2016eik} that the entanglement wedge is the natural bulk region to be associated with a boundary density matrix. Nevertheless, one may wonder if the smaller causal wedge (which is more minimally defined in terms of the bulk causal structure) is not perhaps more fundamental. After all, local bulk operator reconstruction seems to proceed more seamlessly within the causal wedge.\footnote{ This is true with the current technology modulo the subtlety that the HKLL procedure \cite{Hamilton:2006az} involves a nonstandard Cauchy problem.} The fact that the extremal surface allows a decomposition of the bulk spacetime into four distinct domains \cite{Headrick:2014cta}: its future (which we are eschewing in our construction), its past (which is included explicitly), and the entanglement wedges for the region and its complement importantly allows the ability to decompose the bulk semiclassical Hilbert space. This is not possible for the causal wedge (in fact worse still, the causal wedge is not even its own domain of dependence \cite{Hubeny:2014qwa}).

To illustrate this point, consider the eternal black hole viewed as a thermofield double state \cite{Maldacena:2001kr}: if we pick a bulk Cauchy slice that goes through the interior, the modes in the interior do not have a nice interpretation in terms of left or right modes only. A Cauchy surface that passes through the bifurcation surface on the horizon does not encounter this issue.  Generically causal wedges exemplify the former scenario, while entanglement wedges by construction always conform to the latter.\footnote{ While for the eternal black hole the causal wedge and entanglement wedge  for one whole connected component of the boundary coincide, we can deform the black hole using shock-waves (cf., \cite{Headrick:2014cta}) to separate out the causal and entanglement wedges. In the latter scenario, a Cauchy slice is bipartitely divided across the two boundaries by the entanglement wedge but not so by the causal wedge. The causal wedges for the region and its complement fail to meet, being separated by a causal shadow domain.}

\paragraph{Multiple saddles:} While the basic principles of the gravitational problem are general, the final evaluation of the minimal surface condition and obtaining the on-shell action should be viewed as a saddle point analysis. This is valid at 
large central charge: $c_\text{eff} = \frac{\ell_\text{AdS}^{d-1}}{16\pi \, G_N} \gg 1$. Usually when we have multiple solutions to the equations of motion, the saddle point analysis instructs us to pick the one with the least action (which translates here to smallest area for $\extr$). This statement is true for generic R\'enyi index $q$.

It is however important to note that the control parameter for the saddle analysis in the computation of entanglement entropy, i.e., for $q\sim1$, encounters a further suppression by $q-1$. The true parameter is $(q-1) \,c_\text{eff}$. Requiring that this be large as $q\to 1$ is only possible with an appropriate order of limits: we first take $c_\text{eff} \to \infty$ before 
taking $q\to 1$. This point has been noted elsewhere, see for example \cite{Fischetti:2014zja}. 

If one goes to the opposite limit, $(q-1) \,c_\text{eff} \ll 1$, then there is no new saddle for $q \sim 1$, other than the original $q=1$ saddle. Of course, one should not do the saddle point approximation and later set $(q-1) \,c_\text{eff} \ll 1$.  

\paragraph{Lorentzian replicated bulk geometries:}
In the case of local modular Hamiltonians, one can understand explicitly how these Lorentzian geometries look like. 

For example, if we consider a spherical entangling surface in the vacuum, 
\cite{Casini:2011kv} showed that the density matrix was $\rho=e^{-\beta K}|_{\beta=2\pi}$, with $K$ the integral of the stress tensor over $\Sigma_t$. The dual to $D[\regA]$ can be thought as the exterior of the topological black hole. In this way, \cite{Hung:2011nu} explained that the dual of the R\'enyi entropies was given by hyperbolic black hole at inverse temperature $\beta=2\pi q$. Of course, this is a consequence of the density matrix being thermal in the hyperboloid.

However, note that, as explained before, the boundary geometry to compute the R\'enyi entropy should be thought as $q$ geometries glued together. In this way, the dual geometry should look like $q$ asymptotic boundaries (corresponding to $D[\regA]$ in each replica) which are connected together through the interior of the black hole. This is clearly different from a unique black hole. An explicit difference is that while a black hole has four causally disconnected regions (for a $\tau$ observer), the geometry dual to the R\'enyi entropies would have $4 q$ regions.

Of course, the Euclidean picture is the same in the two cases and the R\'enyi entropies coincide. However, they are geometrically different, while changing the temperature is  a boundary at a different temperature: $\tr( \rho_{\scriptscriptstyle{\beta=2\pi q}})$, the R\'enyi entropy is computed by $\tr (\rho_{\scriptscriptstyle{\beta=2\pi}}^q)$. Because the modular Hamiltonian is local the two prescriptions are equivalent and have an explicit dual geometry. In other words, in one case we do Euclidean evolution for $2\pi q$ and then evolve in Lorentzian time, while in the other we have $q$ Euclidean segments where we evolve $2\pi$. At the end of each segment we evolve in Lorentzian time up to some time $t$ and back.

The above is an example of two different  field theory contours that end up having the the same action, since they just differ with each other by some unitaries.  As such one might encounter many contours which end up giving the same on-shell action, but in our discussion we have singled out the  $\mathbb{Z}_q$ symmetric contour which reduces to the canonical Schwinger-Keldysh contour at $q=1$.

\paragraph{Non-analytic geometries:}
  We have exercised care with not extending $\tau$ to the complex domain except in a very particular sense described in \S\ref{sec:grav}. To implement this,  we imposed local Rindler conditions (close to $\fixM$) by expanding the fields in Rindler modes and matching them across the horizons. 

This was inspired by our desire for the construction to apply to generic non-analytic metrics. In this way, even if we define 
the coordinate $\tau$ patchwise, with a discrete imaginary part, there is no analytic continuation involved; the discrete shifts in the imaginary part simply corresponds to a labeling of domains.\footnote{ Recall that this is already useful in the context of Rindler geometry or a black hole spacetime where we can label different domains in the maximally extended Lorentzian spacetime with a discrete imaginary part.}

Nonetheless, it is quite common in gravity, for geometries that satisfy the equations of motion, to be analytic almost everywhere. In such situations one could simply impose the previous conditions by analytically continuing $\tau$ to the entire complex plane and picking an appropriate, continuous contour of integration there. Of course, one could also analytically continue to Euclidean time (which would give us a complex metric in general).

\paragraph{Higher derivative gravitational dynamics:} The main thrust of our analysis has been to derive the HRT proposal for holographic entanglement entropy, which is valid for strongly coupled theories with large $c_\text{eff}$. If we start to include finite coupling $\lambda$ effects then we anticipate that the bulk dynamics is well described by a higher derivative gravitational theory. 

In the time-independent situation a prescription to incorporate such bulk dynamics was given in \cite{Dong:2013qoa,Camps:2013zua}.  The kinematic part of the argument we have presented herein trivially extends to these cases. We then have to work out the local analysis in the vicinity of the fixed point set $\fixM_q$ for the given higher derivative theory.  As in the Euclidean analysis we do not expect that this local analysis will serve to pick out the functional whose variation gives rise to the dynamical constraint on the fixed point set $\fixM_q$; rather one has to work with the full bulk dynamics. However, one can conclude that the functional we should evaluate once we find the surface of interest should be the one obtained in the aforementioned papers.  

As discussed in \cite{Bhattacharjee:2015yaa,Wall:2015raa} the holographic entanglement entropy functionals serve as a good starting point to examine  the second law for higher derivative black hole entropy. The discussion thus far has been confined to the linear response regime of small amplitude fluctuations away from equilibrium. It would be interesting to examine whether one can shed light on the non-linear second law using  some of the machinery developed herein.

\paragraph{Quantum corrections:} A key part of our argument in \S\ref{sec:grav}  was to implement the Schwinger-Keldysh construction directly in the bulk spacetime. This naturally incorporates a  semiclassical separation of Hilbert spaces and thereby allows for a really transparent interpretation of the quantum corrections. Following the arguments of \cite{Faulkner:2013ana}, one can say that the quantum corrections at first subleading order ${\cal O}\left(c_\text{eff}^{-1} \right)$, can be viewed as the entanglement entropy of region $\homsurfA$ in the bulk (suitably regulated). Similar statements can be made for the boundary and bulk relative entropies. As a consequence one can find a purely bulk expression for  boundary modular Hamiltonian as  discussed recently in \cite{Jafferis:2015del}.\footnote{ It is also worth keeping in mind the comments made earlier regarding the entanglement wedge in this context.}  It is useful to recognize that the entanglement wedge naturally implements the Schwinger-Keldysh contour in the bulk; consistent with the fact that the boundary bipartitioning induces a corresponding one in the bulk (cf., the causal domain decomposition discussed in \cite{Headrick:2014cta}).

\paragraph{Maximin construction:} In \cite{Wall:2012uf}, the covariant HRT construction involving extremal surfaces was reformulated as a maximin construction. The primary motivation was a tool in aid to proving strong-subadditivity of holographic entanglement entropy in the time-dependent situations,  extending the initial result of \cite{Headrick:2007km} for static states. 
The idea was to pick a bulk Cauchy slice $\bulkCS$ corresponding to a given region $\regA \subset \CSA$ on the boundary, find a minimal surface on this slice, and then maximize the area of minimal surfaces across a complete set of Cauchy slices inside the Wheeler-DeWitt patch of $\CSA$. 

While the final result of the maximin construction coincides with the extremal surface prescription of HRT, from our point of view not all minimal surfaces on bulk Cauchy slices respect the boundary conditions of the Schwinger-Keldysh construction. The replica construction requires that only the slices that contain $\extr$ are admissible in the bulk path integral. This does not however modify the discussion of strong-subadditivity. All it does is restrict the set of Cauchy slices we need to consider for the maximin construction. 

To wit, one finds a common Cauchy slice $\Sigma$ for two of the spacelike separated regions appearing in the strong-subadditivity inequality (say $\regA_1\cup \regA_2$ and $\regA_1 \cap \regA_2$) which can be argued to exist \cite{Wall:2012uf} --  so ${\cal E}_{{\cal A}_1 \cap {\cal A}_2} , \; {\cal E}_{{\cal A}_1 \cup {\cal A}_2} \subset \Sigma$. 
One then projects the extremal surfaces for regions $\regA_1$ and $\regA_2$ which a priori lie on some other slice, onto 
$\Sigma$, say as $P{\cal E}_{{\cal A}_1}$ and $P{\cal E}_{{\cal A}_2}$  respectively. The key point is that if the bulk theory satisfies null energy condition then area can only decrease under such projection, so $\text{Area}(P{\cal E}_{{\cal A}_i}) \leq 
\text{Area}({\cal E}_{{\cal A}_i})$. Since now all the surfaces lie on a common Cauchy slice we can employ the local surgery argument of \cite{Headrick:2007km} to learn that 
\begin{align}
\text{Area}\left({\cal E}_{{\cal A}_1 \cap {\cal A}_2} \right) + \text{Area} \left({\cal E}_{{\cal A}_1 \cap {\cal A}_2}\right)
\leq \sum_{i=1}^2 \, \text{Area}(P{\cal E}_{{\cal A}_i}) \leq \sum_{i=1}^2 \text{Area}({\cal E}_{{\cal A}_i}) \,, 
\label{}
\end{align}
which establishes the strong-subadditivity result as desired.

\paragraph{Complex saddles:} It has been suggested that complex extremal surfaces could potentially play a role in the computation of holographic entanglement \cite{Fischetti:2014zja,Fischetti:2014uxa}. The argument relies on the fact that typical saddle point evaluations do often admit complex saddles; the derivation of LM \cite{Lewkowycz:2013nqa} could be interpreted in this manner. While the examples discussed in the aforementioned papers are interesting, to our knowledge there is no clear boundary interpretation of the $q=1$ geometries.  It is worth pointing out that in situations where we have some explicit boundary field theory understanding such complex surfaces, while seemingly present, break the time reflection symmetry and appear to be sub-dominant to real saddle points \cite{Maxfield:2014kra}. 

Our take on the problem is rather different. In field theory we are instructed to perform a real time computation
for $\rho(t)$ and thence to manipulate it to construct the reduced density matrix elements; cf., \S\ref{sec:qft}.
As described in detail, the construction extends naturally into the bulk where we glue pieces of the state $\ket{\psi}$
and its conjugate $\bra{\!\psi}$ across some bulk Cauchy slice $\bulkCS$, which contains the extremal surface $\extr$. 
In other words, the computation is phrased purely in Lorentzian terms and leaves no room for purely Euclidean surfaces. 

The skeptical reader may argue that in situations with time reflection symmetry, we could alternately use the Euclidean formulation of the problem as in \S\ref{sec:lmreview}; that computation could potentially be dominated by Euclidean surfaces which do not analytically continue to real Lorentzian saddles. This viewpoint however obscures the temporal ordering inherent in the construction of $\rhoA(t)$ leading to an inherent tension with causality, which is not obviously resolved.  While this issue deserves further investigation let us record here that in general it remains unclear that potential time-reflection ${\mathbb Z}_2$ breaking saddles arise from the $q\to 1$ limit in a replica construction. It would also be useful to ascertain whether there are examples analogous to the ones discussed in \cite{Fischetti:2014uxa} (i.e., absence of real HRT extremal surfaces)  in geometries with known unitary CFT duals.  

%~~~~~~~~~~~~~~~~~~~~~~~~~~~~~~~~~~~~~~~~~~~~~~
\acknowledgments 

%~~~~~~~~~~~~~~~~~~~~~~~~~~~~~~~~~~~~~~~~~~~~~~

It is a pleasure to thank Felix Haehl, Veronika Hubeny, R. Loganayagam, Juan Maldacena, Don Marolf, Henry Maxfield,  Rob Myers, Tadashi Takayanagi, Mark Van Raamsdonk for very useful discussions on various related issues. 

 XD and MR would like to thank the Aspen Center for Physics and KITP, Santa Barbara, for hospitality during the course of this project, where their stays were supported in part by the National Science Foundation (NSF) under grants PHYS-106629 and PHY11-25915 respectively. AL and MR would like to thank the Yukawa Institute for Theoretical Physics, Kyoto for hospitality during the concluding stages of this project. We would also like to thank the Perimeter Institute for Theoretical Physics. Research at Perimeter Institute is supported by the Government of Canada through the Department of Innovation, Science and Economic Development and by the Province of Ontario through the Ministry of Research and Innovation.

 XD was supported in part by the NSF under Grant No. PHY-1316699, by the Department of Energy under Grant No. DE-SC0009988, and by a Zurich Financial Services Membership at the Institute for Advanced Study.  AL was supported in part by the US NSF under Grant No. PHY-1314198.  
%~~~~~~~~~~~~~~~~~~~~~~~~~~~~~~~~~~~~~~~~~~~~~~

\appendix

%~~~~~~~~~~~~~~~~~~~~~~~~~~~~~~~~~~~~~~~~~~~~~~~
\section{Bulk evaluation of the R\'enyi entropy}
\label{sec:action}
%~~~~~~~~~~~~~~~~~~~~~~~~~~~~~~~~~~~~~~~~~~~~~~

For integer $q$,  we will have a well-defined (smooth) action: 
\begin{equation}
I[{\bulk}_q]=\int_{\bulk_q} {\cal L}+\int_{\partial \bulk_q} {\cal L}_{bdy} + \int_{\tilde{\Sigma}_t}\, {\cal L}_{_{SK}}
\end{equation}
 with $\partial \bulk_q$ simply the holographic boundary and the additional boundary term for the Schwinger-Keldysh construction across the codimension-1 bulk Cauchy slice is explicitly included.\footnote{ We are going to assume that at finite $q$ there are no relevant contributions from bulk singularities (if any).} We can use the $\Zn$ symmetry to think about this as $q I[\hat \bulk_q]$, which will again be a local integral in the bulk. 

One should be able to evaluate in principle these partition functions. To do this, one first has to look for the solutions to the equations of motion with the boundary condition \eqref{eq:lmL} for the quotient spacetime $\hat \bulk_q$. These solutions are completely real in the Rindler wedges, but, for even $q$, they might present some imaginary phases in the Milne wedges.\footnote{ Using the coordinates in \eqref{eq:lmL}, the asymptotic boundary tells us that the Milne wedges are reached by $r \rightarrow i^{-1} r$, $\tau \rightarrow \tau + i \frac{\pi}{2}$ and thus those components of the metric with $r^q e^{\pm\tau}$ might get nontrivial phases when $q$ is even.} Once one has a solution, after regularizing the bulk properly, one should be able to compute the action. This procedure was carried out explicitly in $d=2$ from the Euclidean perspective in \cite{Hung:2011nu,Faulkner:2013yia}, although the explicit evaluation of this action is far from being trivial because (among other things) of the regularization of the bulk. While we believe that a similar strategy ought to work in the Lorentzian case, we will not attempt to implement such in the present work.

We expect that if one is careful with how the boundary is regularized and exploits correctly symmetries of the problem, one can get the R\'enyi entropies directly by integrating the action. Due to the presence of light-like singularities, one has to be extremely careful with how they are integrated. From the timefold perspective, we expect that one can recover that the only contribution will come from these singularities, whereas the purely real contribution will cancel. Said differently we expect in the construction a complex action $S = S_R + i\, S_I$ with the timefold ensuring the the pure phase $S_R$ canceling out.\footnote{ We remind the reader that we are working in Lorentz signature and so the action is $e^{i\, S}$.}

As an example, we consider the case of Appendix A of \cite{Lewkowycz:2013nqa}, but for arbitrary Lorentzian sources. 
We describe this explicitly in \S\ref{sec:lmscalar} once we have shown how to carry out the evaluation of the boundary contribution to the on-shell action.

%~~~~~~~~~~~~~~~~~~~~~~~~~~~~~~~~~~~~~~~~~~~~~~~
\subsection{Evaluation of boundary term}
%~~~~~~~~~~~~~~~~~~~~~~~~~~~~~~~~~~~~~~~~~~~~~~

Let us first evaluate  the boundary terms. The contributions from the $r=\epsilon$ pieces of $\fixM_q(\epsilon)$ are simple to evaluate, but they cancel pairwise because of the timefold. So we are left with evaluating the contribution from the jump across the horizon, which we call $\fixM_q^{jump}(\epsilon)$. In the $\epsilon \rightarrow 0$ limit, the boundary is becoming vanishingly small, but it nevertheless leads to a non-vanishing contribution owing to the boundary term which correspondingly diverges. 

As a warm-up, let us consider the two-dimensional Rindler space: $dr^2-r^2 d \tau^2=dx^2-dt^2$, with the codimension-1 surface surrounding the point $r=x=t=0$. The segment $\fixM_q^{jump}(\epsilon)$ joins a spacelike and a timelike surface. The extrinsic curvature will blow up when this segment becomes lightlike. We are going to choose it so that this only happens at a point, and we will employ an $i \varepsilon$ prescription when integrating the boundary term added, not sure of how to say it, because our boundary is not actually physical.\footnote{ The choice of a positive $i \varepsilon$ sign guarantees that if this boundary was physical, the path integral does not blow up. } One can show that this contribution does not depend on the shape of segment and for some initial and final normal vectors $n_1$, $n_2$ it gives \cite{Farhi:1989yr,Neiman:2013ap}:
\begin{equation}
 I_{bdy}= \frac{1}{8\pi\, G_N}\; \int_{\fixM_1^{jump}(\epsilon)} {\cal K}_{\epsilon}=\cosh^{-1}(n_1 \cdot n_2)
\end{equation}
In our case, if we have an initial normal vector $n_1=(\cosh \tau_0,\sinh \tau_0)$, the final normal vector will just be $n_2=i ( \sinh \tau_0,\cosh \tau_0)$.\footnote{ In order to keep the integrand fixed across the whole surface (without changing the determinant of the metric as it flips signature), we have to normalize the normal vectors to $1$.}  We can also see this by thinking of crossing the horizon as shifting $\tau_0 \rightarrow \tau_0+i \frac{\pi}{2}$. 
For $q=1$ we cross the horizon four times and get $4 I_{bdy}=i \frac{\text{Area}(\fixM_1)}{4 G_N} $.

When we have the quotient space metric \eqref{eq:lmL}: $q^2 dr^2-r^2 d\tau^2+\gamma_{ij} dy^i dy^j+ \cdots$, it is a little more subtle how to apply the previous argument. In order to compare the normal vectors one has to go to locally flat coordinates: $x=r \cosh \tau/q$, $t=r \sinh \tau/q$ and compare the normal vectors in these coordinates. These coordinates are not really physical, but they are useful to compare the normals. As we cross the horizon $r \rightarrow i^{-1} r$ and $\tau \rightarrow \tau+i \pi/2$, so if the initial normal is $n_1=(\cosh \tau_0/q,\sinh \tau_0/q)$ then $n_2=(\cosh \frac{\tau_0+i \pi/2}{q},\sinh \frac{\tau_0+i \pi/2}{q})$  and $n_1 \cdot n_2=\cosh \frac{i \pi}{2 q}$, so
\begin{equation}
I_{bdy}[\hat{\bulk}_q]=4 \frac{i}{4 q} \frac{\text{Area}(\fixM_q)}{4 \, G_N} \,.
\end{equation}

 ~~~~~~~~~~~~~~~~~~~~~~~~~~~~~~~~~~~~~~~~~~~~~
\subsection{Example of bulk integral}
\label{sec:lmscalar}
~~~~~~~~~~~~~~~~~~~~~~~~~~~~~~~~~~~~~~~~~~~~~

The contribution to the entropy of one interval in the vacuum from a  time-dependent scalar source was computed in \cite{Lewkowycz:2013nqa} (see their Appendix A). To do this, we first consider a bulk scalar $\phi$ with boundary condition $\phi|_{\partial \bulk_q}=\lambda \varphi(\tau)$ and $\lambda \ll 1$, and then compute the $O(\lambda^2)$ contribution to the R\'enyi entropy.  The action will be 
\begin{equation}
{I}[{\hat{\bulk}_q}]={I}_{EH}[{\hat{\bulk}_q}] - \int_{\hat{\bulk}_q} d^{d+1}x \sqrt{-g} 
\left[  (\partial \phi)^2+V(\phi) \right]
\end{equation}

In the absence of the scalar term, the calculation will be that of \cite{Hung:2011nu}.  
The argument from the main text tells us that the $O(\lambda^2)$ contribution of $\tilde{S}_\regA^{(q)}$ will be given by the area of the replica fixed point after accounting for the leading backreaction of the scalar field, but it was also shown in \cite{Lewkowycz:2013nqa} that this is equivalent to the purely matter contribution for the modular entropy:\footnote{ We refer the reader to Appendix A of \cite{Lewkowycz:2013nqa}  for more details.} 
\begin{equation}
\tilde{S}_\regA^{(q)}|_{\lambda^2}=\frac{A}{4 q^2 \,G_N}\bigg|_{\lambda^2}
=-i\int_{\hat{\bulk}_q} T^{\phi}_{\mu \nu} \; \partial_q g^{\mu \nu}
=-i\partial_q \int_{\partial \hat{\bulk}_q}\phi\, \partial_{\vec{n}} \phi 
\label{eq:matterentropy}
\end{equation}
where $\partial_{\vec n}$ denotes the derivative normal to the surface. Even though the area term is purely gravitational, the other two denote different expressions for the variation of the area purely in terms of the scalar solutions. While the integral looks imaginary, the presence of a light-cone singularity in the integrand implies that we need to implement a proper $i\varepsilon$ prescription in evaluating the integral. The result, as we shall see, will be a  real answer for $\tilde{S}_\regA^{(q)}|_{\lambda^2}$ in \eqref{eq:matterentropy}.

Since the replicated solutions are complicated, we set $q=1$ in (\ref{eq:matterentropy}) from now on. We would like to generalize this argument to the Lorentzian case, by inserting general time-dependent sources (not only time-dependent in the $\tau$ direction as in \cite{Lewkowycz:2013nqa}). In other words, we want to consider a scalar profile that depends on $t$ in the Poincar\'e patch:
\begin{eqnarray}
ds^2=\frac{dx^2-dt^2+dz^2}{z^2} \,.
\end{eqnarray}
However, in order to compute $\partial_q g$, it is easier to work in the hyperbolic patch:
\begin{eqnarray}
ds^2=f(\rho) \frac{du^2}{u^2}-\rho^2 d\tau^2+\frac{d\rho^2}{f(\rho)} \,.
\end{eqnarray}
These two geometries correspond to $q=1$ (with no backreaction from the scalar), with $f(\rho)=\rho^2+1$, and the two coordinate systems are related by $\rho^2=\dfrac{x^2-t^2}{z^2}$, $u^2=x^2-t^2+z^2$, and $\tanh \tau=\frac{t}{x}$. In hyperbolic coordinates, the replicated geometry is just $f_q(\rho)=\rho^2+q^{-2}$.

For technical reasons, it is easy to compute R\'enyi entropies in hyperbolic coordinates, but difficult in Poincar\'e coordinates. This means that if we want to compute the contribution from the scalar action \eqref{eq:matterentropy}, this will be particularly simple when evaluating the bulk stress tensor contribution at $q=1$. We are going to focus on this case, i.e., we are going to compute $\partial_q g$ using the hyperbolic coordinates and then go to $q=1$ where we have the explicit hyperbolic-to-Poincar\'e coordinate transformation. If one wanted to compute the \emph{boundary} matter contribution from $\partial \bulk_q$ in \eqref{eq:matterentropy}, one would have to be careful with the regularization of the manifold close to the boundary; we will leave this for the future.

Even though the evaluation of $\tilde{S}_\regA^{(1)}|_{\lambda^2}=-i\int T_{\mu \nu} \partial_q g^{\mu \nu}$ has to trivially give us the area (because of the argument of \cite{Lewkowycz:2013nqa}), we believe that this is a simple example of how the general case works. More concretely, we just need to show that, if we analytically continue (i.e., $t_E\rightarrow i t)$ the Euclidean calculation with the proper prescription, the answer is $i$ times the Euclidean one. Then the analysis of \cite{Lewkowycz:2013nqa} tells us that we get the area. 

Let us consider a scalar solution $\lambda e^{i \omega t} f_{\omega}(z)$.\footnote{ Note that $f(z)$ solves the scalar equations of motion, but we will not need its explicit form.} By expressing everything in Poincar\'e coordinates, we obtain:
\begin{equation}
\begin{split}
i  \tilde{S}_\regA^{(1)}|_{\lambda^2} &= \int_{{\cal M}} \tilde{T}^{\phi}\,,
\\
 \tilde{T}^{\phi} &\equiv \sqrt{-g} \;T_{\mu \nu}^{\phi} \;\partial_q g^{\mu \nu}  \\
 &= \frac{2 z}{(l^2+z^2)^2} \left [ (T^{\phi}_{l l}-T^{\phi}_{zz})  (l^2-z^2)+4\, l \,z \,T^{\phi}_{z l}\right ]  \\
& = \frac{2\, z \,(l^2-z^2)}{\left(l^2+z^2\right)^2} \( \frac{t^2}{l^2} \,\omega^2\, f_{\omega} f_{-\omega} -f_{\omega}' f_{-\omega}'\)+\frac{4 \,i \,t \,z^2 \,\omega}{(l^2+z^2)^2} (f_{\omega} f_{-\omega}'-f_{-\omega} f'_{\omega}) 
\end{split}
\end{equation}
 where for simplicity we have used $l^2=x^2-t^2$, but we want to think about the previous expression in terms of $x,t,z$.  

 Now, the question becomes a very simple one. We can integrate $\tilde{T}^\phi$ in Euclidean signature with respect to $\tE=i t$ and leave it as a function of $x,z$; since there are no poles, this is straightforward. The argument of Appendix A of \cite{Lewkowycz:2013nqa} shows that $\int_{{\cal M}} \tilde{T}^\phi(\tE)=\frac{A}{4 G_N}$.   We only have to check that the previous Lorentzian integral gives $i$ times the Euclidean integral. One can see this as follows: first integrate the previous expression with respect to $t$. This will give a result involving just the integrals over $x$ and $z$, which is insensitive to the Lorentzian or Euclidean signature. The integral over $t$ involves poles at $t^2 = x^2 + z^2$ in the Lorentzian signature, which we perform  by choosing an $i \varepsilon$ prescription.  This leads to the desired factor of $i$, and the final integrand over $x$ and $z$ is $i$ times the Euclidean one.

%%%%%%%%%%%%%%%%%%%%%%%%%%%%%%%%%%%%%%%%%%%%
% \bibliographystyle{jhep}
% \bibliography{entanglement}

\providecommand{\href}[2]{#2}\begingroup\raggedright\endgroup

\end{document}